\documentclass[11pt]{article}
\usepackage{amssymb,amsmath,amsfonts}
\usepackage{graphicx}
\usepackage{graphics}
\usepackage{eepic,epsfig}

\@addtoreset{equation}{section}

\makeatother

\begin{document}

\title{Fermionic vacuum polarization by a cosmic string\\
in de Sitter spacetime }
\author{E. R. Bezerra de Mello$^{1}$\thanks{%
E-mail: emello@fisica.ufpb.br}, \thinspace\ A. A. Saharian$^{1,2}$\thanks{%
E-mail: saharian@ysu.am} \\
\textit{$^{1}$Departamento de F\'{\i}sica, Universidade Federal da Para\'{\i}%
ba}\\
\textit{58.059-970, Caixa Postal 5.008, Jo\~{a}o Pessoa, PB, Brazil}\vspace{%
0.3cm}\\
\textit{$^2$Department of Physics, Yerevan State University,}\\
\textit{1 Alex Manoogian Street, 0025 Yerevan, Armenia}}
\maketitle

\begin{abstract}
We investigate the fermionic condensate and the vacuum expectation value of
the energy-momentum tensor for a massive spinor field in the geometry of a
straight cosmic string on background of de Sitter spacetime. By using the
Abel-Plana summation formula, we explicitly extract form the expectation
values the contribution associated with purely de Sitter space, remaining
the expectation values induced by the cosmic string. The latter presents
information about de Sitter gravity as well. Because the investigation of
the fermionic quantum fluctuations in de Sitter space have been investigated
in literature, here we are mainly interested in the cosmic string-induced
contributions. For a massless field, the fermionic condensate vanishes and
the presence of the string does not break chiral symmetry of the massless
theory. Unlike to the case of a scalar field, for a massive fermionic field
the vacuum expectation value of the energy-momentum tensor is diagonal and
the axial and radial stresses are equal to the energy density. At large
distances from the string the behavior of the string-induced parts in the
vacuum densities is damping oscillatory with the amplitude decaying as the
inverse fourth power of the distance. This is in contrast to the case of
flat spacetime, in which the string-induced vacuum densities for a massive
field decay exponentially with distance from the string. In the limit of the
large curvature radius of de Sitter space we recover the results for a
cosmic string in flat spacetime.
\end{abstract}

\bigskip


\bigskip

\section{Introduction}

Different types of topological objects may have been formed in the
early universe after Planck time by the vacuum phase transition
\cite{Vile94}. Depending on the topology of the vacuum manifold
these are domain walls, strings, monopoles and textures. Among
them the cosmic strings are of special interest. Although the
recent observational data on the cosmic microwave background
radiation have ruled out cosmic strings as the primary source for
primordial density perturbations, they are still candidates for
the generation of a number of interesting physical effects such as
gravitational lensing, cosmic microwave background
non-gaussianities, the emission of gravitational waves and
high-energy cosmic rays (see, for instance, \cite{Damo00}). More
recently, cosmic strings attract a renewed interest partly because
a variant of their formation mechanism is proposed in the
framework of brane inflation \cite{Sara02}.

In the simplest theoretical model describing the infinite straight cosmic
string the spacetime is locally flat except on the string where it has a
delta shaped Riemann curvature tensor. In quantum field theory the
corresponding non-trivial topology induces non-zero vacuum expectation
values for physical observables. Explicit calculations for the geometry of a
single cosmic string have been done for different fields \cite{Hell86}-\cite%
{Spin08}. Vacuum polarization effects by higher-dimensional composite
topological defects constituted by a cosmic string and global monopole are
investigated in Refs. \cite{Beze06Comp} for scalar and fermionic fields.
Another type of vacuum polarization arises when boundaries are present. The
imposed boundary conditions on quantum fields alter the zero-point
fluctuations spectrum and result in additional shifts in the vacuum
expectation values of physical quantities. In Ref. \cite{Beze06sc}, we have
studied both types of sources for the polarization of the scalar vacuum,
namely, a cylindrical boundary and a cosmic string, assuming that the
boundary is coaxial with the string and that on this surface the scalar
field obeys Robin boundary condition. The polarization of the
electromagnetic vacuum by a conducting cylindrical shell in the cosmic
string spacetime is investigated in \cite{Beze07El}. The case of the
fermionic field with bag boundary condition on the cylindrical surface is
discussed in \cite{Beze08}.

Many of treatments of quantum fields around a cosmic string deal mainly with
the flat background geometry. Quantum effects for a scalar field produced by
a string on curved backgrounds have been investigated in \cite{Davi88} for
special values of planar angle deficit when the corresponding two-point
functions can be constructed by making use of the method of images. In a
previous paper \cite{Beze09} we have investigated the vacuum polarization
effect associated with a quantum massive scalar field in de Sitter (dS)
spacetime in the presence of a cosmic string (for cosmic strings in
background of dS spacetime see \cite{Line86}-\cite{Beze03}). It has been
shown that the corresponding gravitational field essentially changes the
behavior of the vacuum densities at distances from the string larger than
the dS curvature radius. Depending on the curvature radius of de Sitter
spacetime, two regimes are realized with monotonic and oscillatory behavior
of the vacuum expectation values at large distances. Another interesting
feature due to the background gravitational field is the appearance of
non-zero off-diagonal component of the energy-momentum tensor which
corresponds to the energy flux along the radial direction (see, also, \cite%
{Davi88}; a similar effect induced by plane boundaries in dS spacetime has
been observed in \cite{Saha09}). In \cite{Saha10} we have considered the
influence of a cosmic string to the power-spectrum of quantum fluctuations
for a scalar field in dS spacetime. The dependence of the power-spectrum on
the distance from the string is oscillatory with the period having the order
of the wavelength for the perturbation. One-loop topological quantum effects
for scalar and fermionic fields induced by the toroidal compactification of
spatial dimensions in dS spacetime have been recently considered in \cite%
{Saha08dS,Saha08dSf}.

In the present paper we provide the results of the investigation for the
fermionic vacuum polarization by a cosmic string in dS spacetime. dS
spacetime is the maximally symmetric solution of the Einstein equations with
a positive cosmological constant and due to its high symmetry numerous
physical problems are exactly solvable on this background. A better
understanding of physical effects in this background could serve as a handle
to deal with more complicated geometries. In most inflationary models an
approximately dS spacetime is employed to solve a number of problems in
standard cosmology \cite{Lind90}. More recently astronomical observations of
high redshift supernovae, galaxy clusters and cosmic microwave background
\cite{Ries07} indicate that at the present epoch the universe is
accelerating and can be well approximated by a world with a positive
cosmological constant. If the universe would accelerate indefinitely, the
standard cosmology would lead to an asymptotic dS universe. In addition to
the above, an interesting topic which has received increasing attention is
related to string-theoretical models of dS spacetime and inflation. Recently
a number of constructions of metastable dS vacua within the framework of
string theory are discussed (see, for instance, \cite{Kach03} and references
therein).

The results obtained here can be used, in particular, for the investigation
of the effects of the quantum fluctuations induced by the string in the
inflationary phase. Though the cosmic strings produced in phase transitions
before or during early stages of inflation would have been drastically
diluted by the expansion, the formation of defects during inflation can be
triggered by a coupling of the symmetry breaking field to the inflaton field
or to the curvature of the background spacetime (see \cite{Vile94}). Cosmic
strings can also be continuously created during inflation by
quantum-mechanical tunnelling \cite{Basu91}. Another class of models to
which the results of the present paper are applicable corresponds to
string-driven inflation where the cosmological expansion is driven entirely
by the string energy \cite{Turo88}. The problem under consideration is also
of separate interest as an example with gravitational and topological
polarizations of the fermionic vacuum, where all calculations can be
performed in a closed form.

The plan of the paper is the following. In the next section, the geometry
under consideration is described and a complete set of solutions to Dirac
equation is constructed. In section \ref{sec:Cond} we evaluate the fermionic
condensate by using the mode-summation method. The string induced part is
explicitly extracted and its behavior in the asymptotic regions of the
parameters are investigated. The vacuum expectation value of the
energy-momentum tensor is considered in section \ref{sec:EMT}. The main
results of the paper are summarized in section \ref{sec:Conc}. The Appendix %
\ref{sec:App1} contains some technical details of the obtainment of the
vacuum expectation values. In Appendix \ref{sec:App2} we consider the
expectation values in dS spacetime when the string is absent.

\section{Fermionic eigenfunctions}

\label{sec:EigFunc}

The main objective of the present section is to obtain the complete set of
solutions of Dirac equation in dS spacetime in presence of an infinitely
long straight cosmic string. In order to do that, we write the corresponding
line element in cylindrical coordinates, having the linear defect along the $%
z$-axis:%
\begin{equation}
ds^{2}=g_{\mu \nu }dx^{\mu }dx^{\nu }=dt^{2}-e^{2t/\alpha
}(dr^{2}+r^{2}d\phi ^{2}+dz{}^{2}),  \label{ds21}
\end{equation}%
where $r\geqslant 0$, $-\infty <z<\infty $, $0\leqslant \phi \leqslant \phi
_{0}\leqslant 2\pi $ and the spatial points $(r,\phi ,z)$ and $(r,\phi +\phi
_{0},z)$ are to be identified. The parameter $\alpha $ in (\ref{ds21}) is
related with the cosmological constant $\Lambda $ by the formula $\alpha =%
\sqrt{3/\Lambda }$. Making use of the coordinate transformation%
\begin{eqnarray}
t &=&t_{s}-\alpha \ln f(r_{s}),\;r=r_{s}f(r_{s})e^{-t_{s}/\alpha }\sin
\theta ,  \notag \\
\;z &=&r_{s}f(r_{s})e^{-t_{s}/\alpha }\cos \theta ,\;\phi =\phi ,
\label{Coord}
\end{eqnarray}%
with $f(r_{s})=1/\sqrt{1-r_{s}^{2}/\alpha ^{2}}$, the line element (\ref%
{ds21}) is written in static form
\begin{equation}
ds^{2}=f^{-2}(r_{s})dt_{s}^{2}-f^{2}(r_{s})dr_{s}^{2}-r_{s}^{2}(d\theta
^{2}+\sin ^{2}\theta d\phi ^{2}).  \label{dSstatic}
\end{equation}%
Rescaling the angular variable in accordance with $\varphi =2\pi \phi /\phi
_{0}$, we transform the metric corresponding to (\ref{dSstatic}) into the
form previously discussed in \cite{Ghez02}. In this paper it is shown that
to leading order in the gravitational coupling the effect of the vortex on
de Sitter spacetime is to create a deficit angle in the metric (\ref%
{dSstatic}).

The dynamics of a massive spinor field on a curved spacetime are described
by Dirac equation
\begin{equation}
i\gamma ^{\mu }\nabla _{\mu }\psi -m\psi =0\ ,\;\nabla _{\mu }=\partial
_{\mu }+\Gamma _{\mu },  \label{Direq}
\end{equation}%
where $\gamma ^{\mu }$ are the Dirac matrices in curved spacetime and $%
\Gamma _{\mu }$ is the spin connection. They are given in terms of the flat
space Dirac matrices $\gamma ^{(a)}$ by the relations
\begin{equation}
\gamma ^{\mu }=e_{(a)}^{\mu }\gamma ^{(a)},\;\Gamma _{\mu }=\frac{1}{4}%
\gamma ^{(a)}\gamma ^{(b)}e_{(a)}^{\nu }e_{(b)\nu ;\mu }\ ,  \label{Gammamu}
\end{equation}%
where the semicolon means the standard covariant derivative for vector
fields. In (\ref{Gammamu}), $e_{(a)}^{\mu }$ is the tetrad field satisfying
the relation $e_{(a)}^{\mu }e_{(b)}^{\nu }\eta ^{ab}=g^{\mu \nu }$, with $%
\eta ^{ab}$ being the Minkowski spacetime metric tensor.

In the discussion below the flat space Dirac matrices will be taken in the
standard form \cite{Bjor64}%
\begin{equation}
\gamma ^{(0)}=\left(
\begin{array}{cc}
1 & 0 \\
0 & -1%
\end{array}%
\right) ,\;\gamma ^{(a)}=\left(
\begin{array}{cc}
0 & \sigma _{a} \\
-\sigma _{a} & 0%
\end{array}%
\right) ,\;a=1,2,3,  \label{gam0l}
\end{equation}%
with $\sigma _{1},\sigma _{2},\sigma _{3}$ being the Pauli matrices. The
tetrad fields corresponding to line element (\ref{ds21}) may have the form%
\begin{equation}
e_{(a)}^{\mu }=e^{-t/\alpha }\left(
\begin{array}{cccc}
e^{t/\alpha } & 0 & 0 & 0 \\
0 & \cos (q\phi ) & -\sin (q\phi )/r & 0 \\
0 & \sin (q\phi ) & \cos (q\phi )/r & 0 \\
0 & 0 & 0 & 1%
\end{array}%
\right) ,  \label{Tetrad}
\end{equation}%
where
\begin{equation}
q=2\pi /\phi _{0}\geqslant 1.  \label{qu}
\end{equation}%
For the curved space gamma matrices this leads to the formulae%
\begin{equation}
\gamma ^{0}=\gamma ^{(0)},\;\gamma ^{l}=e^{-t/\alpha }\left(
\begin{array}{cc}
0 & \beta ^{l} \\
-\beta ^{l} & 0%
\end{array}%
\right) ,  \label{gaml}
\end{equation}%
with the $2\times 2$ matrices%
\begin{equation}
\beta ^{1}=\left(
\begin{array}{cc}
0 & e^{-iq\phi } \\
e^{iq\phi } & 0%
\end{array}%
\right) ,\;\beta ^{2}=-\frac{i}{r}\left(
\begin{array}{cc}
0 & e^{-iq\phi } \\
-e^{iq\phi } & 0%
\end{array}%
\right) ,\;\;\beta ^{3}=\left(
\begin{array}{cc}
1 & 0 \\
0 & -1%
\end{array}%
\right) .  \label{betal}
\end{equation}%
In (\ref{gaml}) and below the index $l$ runs over values $1,2,3$. For the
components of the spin connection \ we find%
\begin{equation}
\Gamma _{0}=0,\;\Gamma _{l}=-\frac{1}{2\alpha }\gamma ^{0}\gamma _{l}+\frac{%
1-q}{2}\gamma ^{(1)}\gamma ^{(2)}\delta _{l}^{2}.  \label{Gaml}
\end{equation}%
This leads to the following expression for the combination appearing in
Dirac equation:
\begin{equation}
\gamma ^{\mu }\Gamma _{\mu }=\frac{3\gamma ^{0}}{2\alpha }+\frac{1-q}{2r}%
\gamma ^{1}.  \label{gamGam}
\end{equation}%
For further analysis we shall adopt the conformal time $\tau $ defined by $%
\tau =-\alpha e^{t/\alpha }$, $-\infty <\tau <0$.

Decomposing the bispinor $\psi $ into the upper and lower two-component
spinors, denoted by $\varphi $ and $\chi $ respectively, Dirac equation is
written in the form of two coupled first order differential equations:
\begin{eqnarray}
\hat{D}_{+}\varphi +\left( \beta ^{l}\partial _{l}\mathbf{+}\frac{1-q}{2r}%
\beta ^{1}\right) \chi  &=&0,  \label{DirEq1} \\
\hat{D}_{-}\chi +\left( \beta ^{l}\partial _{l}\mathbf{+}\frac{1-q}{2r}\beta
^{1}\right) \varphi  &=&0.  \label{DirEq2}
\end{eqnarray}%
Here we have defined the operators%
\begin{equation}
\hat{D}_{\pm }=\partial _{\tau }-\frac{1}{\tau }\left( \frac{3}{2}\pm
im\alpha \right) .  \label{Dtau}
\end{equation}%
Note that we are still working in the coordinate system corresponding to
line element (\ref{ds21}).

By using the properties of the matrices (\ref{betal}), we obtain the
following second order differential equation for the upper component of the
bispinor:
\begin{equation}
\left[ \partial _{r}^{2}+\frac{1}{r}\partial _{r}+\frac{1}{r^{2}}\partial
_{\phi }^{2}+\partial _{z}^{2}+\frac{q-1}{r}\beta ^{1}\beta ^{2}\partial
_{\phi }\mathbf{-}\frac{(q-1)^{2}}{4r^{2}}-\hat{D}_{+}\hat{D}_{-}\right]
\varphi =0.  \label{Eqphi}
\end{equation}%
Similar equation is obtained for the lower component $\chi $ with $\hat{D}%
_{-}\hat{D}_{+}$ instead of $\hat{D}_{+}\hat{D}_{-}$. Since the matrix $%
\beta ^{1}\beta ^{2}$ is diagonal, from equation (\ref{Eqphi}) it follows
that if we write
\begin{equation}
\varphi =\left(
\begin{array}{c}
\varphi ^{(1)} \\
\varphi ^{(2)}%
\end{array}%
\right) ,  \label{phidecomp}
\end{equation}%
then the equations for the upper and lower components are decomposed:%
\begin{equation}
\left[ \partial _{r}^{2}+\frac{1}{r}\partial _{r}+\frac{1}{r^{2}}\partial
_{\phi }^{2}+\partial _{z}^{2}-(-1)^{a}i\frac{q-1}{r^{2}}\partial _{\phi }%
\mathbf{-}\frac{(q-1)^{2}}{4r^{2}}-\hat{D}_{+}\hat{D}_{-}\right] \varphi
^{(a)}=0.  \label{phijEq}
\end{equation}%
with $a=1,2$.

The solution of equation (\ref{phijEq}) can be presented in the form%
\begin{equation}
\varphi ^{(a)}=f^{(a)}(\tau ,r)e^{i\left( qn_{a}\phi +kz\right) },
\label{phia}
\end{equation}%
with $n_{a}=0,\pm 1,\pm 2,\ldots $, $-\infty <k<\infty $, and with the
equation for the function $f^{(a)}(\tau ,r)$:
\begin{equation}
\left( \partial _{r}^{2}\mathbf{+}\frac{1}{r}\partial _{r}-\frac{\beta
_{a}^{2}}{r^{2}}-k^{2}-\hat{D}_{+}\hat{D}_{-}\right) f^{(a)}(\tau ,r)=0.
\label{fjEq}
\end{equation}%
Here we have introduced the notation
\begin{equation}
\beta _{a}=|qn_{a}-(-1)^{a}(q-1)/2|.  \label{betaj}
\end{equation}%
From (\ref{fjEq}) it follows that the time and radial coordinate dependences
of the function $f^{(a)}(\tau ,r)$ can be separated with the solution%
\begin{equation}
f^{(a)}(\tau ,r)=T^{(a)}(\tau )J_{\beta _{a}}(\lambda r),  \label{fjdec}
\end{equation}%
where $0\leqslant \lambda <\infty $ and $J_{\nu }(x)$ is the Bessel function.

For the function $T^{(a)}(\tau )$ in (\ref{fjdec}) one finds the equation%
\begin{equation}
T^{(a)\prime \prime }(\tau )-\frac{3}{\tau }T^{(a)\prime }(\tau )+\left(
\lambda ^{2}+k^{2}+\frac{m^{2}\alpha ^{2}+im\alpha +15/4}{\tau ^{2}}\right)
T^{(a)}(\tau )=0.  \label{Teq}
\end{equation}%
The general solution for this equation has the form $T^{(a)}(\tau )=\tau
^{2}T_{a\varphi }(\tau )$ with%
\begin{equation}
T_{a\varphi }(\tau )=\sum_{l=1,2}C_{a\varphi }^{(l)}H_{1/2-im\alpha
}^{(l)}(\gamma \eta ),  \label{Tjphi}
\end{equation}%
where%
\begin{equation}
\eta =|\tau |,\;\gamma =\sqrt{\lambda ^{2}+k^{2}},  \label{etagam}
\end{equation}%
and $H_{\nu }^{(l)}(x)$ are the Hankel functions. Hence, for the upper and
lower components of the spinor $\varphi $ we have the solutions%
\begin{equation}
\varphi ^{(a)}=\eta ^{2}T_{a\varphi }(\tau )J_{\beta a}(\lambda r)e^{i\left(
qn_{a}\phi +kz\right) }.  \label{phijsol}
\end{equation}%
Note that the following relation takes place:%
\begin{equation}
\hat{D}_{+}\left[ \tau ^{2}T_{a\varphi }(\tau )\right] =-\gamma \eta
^{2}\sum_{l=1,2}C_{a\varphi }^{(l)}H_{-1/2-im\alpha }^{(l)}(\gamma \eta ).
\label{D+T}
\end{equation}

In a similar way, it can be seen that if we write the lower component of the
bispinor as%
\begin{equation}
\chi =\left(
\begin{array}{c}
\chi ^{(1)} \\
\chi ^{(2)}%
\end{array}%
\right) ,  \label{xidec}
\end{equation}%
then for the separate functions one has the solution
\begin{equation}
\chi ^{(a)}=\tau ^{2}T_{a\chi }(\tau )J_{\tilde{\beta}_{a}}(\lambda
r)e^{i\left( q\tilde{n}_{a}\phi +kz\right) },\;a=1,2.  \label{xij}
\end{equation}%
Here $\tilde{n}_{a}=0,\pm 1,\pm 2,\ldots $, $\tilde{\beta}_{a}=|q\tilde{n}%
_{a}-(-1)^{a}(q-1)/2|$, and we have defined the functions
\begin{equation}
T_{a\chi }(\tau )=\sum_{l=1,2}C_{a\chi }^{(l)}H_{-1/2-im\alpha
}^{(l)}(\gamma \eta ).  \label{Tjxi}
\end{equation}%
For these functions one has the relations%
\begin{equation}
\hat{D}_{-}\left[ \tau ^{2}T_{a\chi }(\tau )\right] =\gamma \eta
^{2}\sum_{l=1,2}C_{a\chi }^{(l)}H_{1/2-im\alpha }^{(l)}(\gamma \eta ).
\label{D-T}
\end{equation}

Now, from equations (\ref{DirEq1}) and (\ref{DirEq2}) we obtain relations
between the parameters in the solutions for the functions $\varphi ^{(a)}$
and $\chi ^{(a)}$:%
\begin{equation}
n_{2}=n_{1}+1,\;\beta _{2}=\beta _{1}+\epsilon _{n_{1}},\;\tilde{n}%
_{a}=n_{a},\;\tilde{\beta}_{a}=\beta _{a},  \label{reln21}
\end{equation}%
and for the coefficients in the linear combinations of the Hankel functions:%
\begin{eqnarray}
\gamma C_{1\chi }^{(l)} &=&-\lambda \epsilon _{n_{1}}C_{2\varphi
}^{(l)}-ikC_{1\varphi }^{(l)},  \notag \\
\gamma C_{2\chi }^{(l)} &=&\lambda \epsilon _{n_{1}}C_{1\varphi
}^{(l)}+ikC_{2\varphi }^{(l)},  \label{Crel}
\end{eqnarray}%
with $l=1,2$, and $\epsilon _{n}=1$ for $n\geqslant 0$ and $\epsilon _{n}=-1$
for $n<0$.

Different choices of the coefficients in the linear combination (\ref{Tjphi}%
) correspond to different choices of the vacuum state. We will consider a dS
invariant Bunch-Davies vacuum (also known as the euclidean vacuum) \cite%
{Bunc78} for which the coefficient for the part containing the function $%
H_{1/2-im\alpha }^{(2)}(\gamma \eta )$ is zero: $C_{a\varphi }^{(2)}=0$. The
Bunch-Davies vacuum reduces to the standard Minkowski vacuum when dS
curvature radius is taken to infinity. From relations (\ref{Crel}) it
follows that $C_{a\chi }^{(2)}=0$. Note that with this choice, the solution
under consideration reduces to the standard positive frequency solutions in
the limit $\eta \rightarrow \infty $. In this sense, we will refer to the
corresponding solutions in dS spacetime as positive frequency solutions (the
same for the negative frequency solutions, see below). Hence, for the
positive frequency modes we have%
\begin{equation}
\psi _{\sigma }^{(+)}(x)=\frac{\eta ^{2}}{\gamma }C_{1\varphi
}^{(1)}e^{i\left( qn\phi +kz\right) }\left(
\begin{array}{c}
\gamma H_{1/2-im\alpha }^{(1)}(\gamma \eta )J_{\beta _{1}}(\lambda r) \\
\gamma C_{\varphi }^{(1)}H_{1/2-im\alpha }^{(1)}(\gamma \eta )J_{\beta
_{2}}(\lambda r)e^{iq\phi } \\
(\lambda \epsilon _{n}C_{\varphi }^{(1)}+ik)H_{-1/2-im\alpha }^{(1)}(\gamma
\eta )J_{\beta _{1}}(\lambda r) \\
-(\lambda \epsilon _{n}+ikC_{\varphi }^{(1)})H_{-1/2-im\alpha }^{(1)}(\gamma
\eta )J_{\beta _{2}}(\lambda r)e^{iq\phi }%
\end{array}%
\right) ,  \label{psisig+}
\end{equation}%
where $n=0,\pm 1,\pm 2,\ldots $, $C_{\varphi }^{(1)}=C_{2\varphi
}^{(1)}/C_{1\varphi }^{(1)}$, and%
\begin{equation}
\beta _{1}=|q(n+1/2)-1/2|,\;\beta _{2}=\beta _{1}+\epsilon _{n}.
\label{bet12}
\end{equation}%
Here the index $\sigma $ for the eigenfunctions stands for the set of
quantum numbers specifying the solution. This set will be specified below.

For the further specification of the eigenfunctions, following \cite{Skar94}%
, we define the operator
\begin{equation}
\hat{S}=\gamma ^{-1}\Sigma ^{l}\ \hat{p}_{l}\ ,  \label{Sop}
\end{equation}%
with
\begin{equation}
\hat{p}_{1}=-i\partial _{r}\ ,\ \hat{p}_{2}=-i\partial _{\phi }+\frac{q-1}{2}%
\Sigma ^{3},\;\ \hat{p}_{3}=-i\partial _{z}  \label{prop}
\end{equation}%
and
\begin{equation}
\Sigma ^{l}=\left(
\begin{array}{cc}
\beta ^{l} & 0 \\
0 & \beta ^{l}%
\end{array}%
\right) \ .  \label{Sigmal}
\end{equation}%
Imposing that the solutions $\psi _{\sigma }^{(+)}$ obey the condition
\begin{equation}
\hat{S}\psi _{\sigma }^{(+)}=s\psi _{\sigma }^{(+)}\ ,  \label{Spsi}
\end{equation}%
for the eigenvalues $s$ and for the coefficient in (\ref{psisig+}) one finds
\begin{equation}
s=\pm 1,\;C_{\varphi }^{(1)}=\frac{i\lambda \epsilon _{n}}{k+s\gamma }.
\label{sval}
\end{equation}

We can see that with this choice, the bispinor (\ref{psisig+}) is an
eigenfunction for the projection of the total momentum along the cosmic
string:%
\begin{equation}
\widehat{J}_{3}\psi _{\sigma }^{(+)}=\left( -i\partial _{\phi }+i\frac{q}{2}%
\gamma ^{(1)}\gamma ^{(2)}\right) \psi _{\sigma }^{(+)}=qj\psi _{\sigma
}^{(+)},  \label{J3m}
\end{equation}%
where%
\begin{equation}
j=n+1/2,\;j=\pm 1/2,\pm 3/2,\ldots .  \label{jn}
\end{equation}%
Hence, as a set of quantum numbers $\sigma $ specifying the solutions we can
take the set $(\lambda ,k,j,s)$. Note that we can write the expressions for
the orders of the Bessel functions in the form%
\begin{equation}
\beta _{1}=q|j|-\epsilon _{j}/2,\;\beta _{2}=q|j|+\epsilon _{j}/2.
\label{betajn}
\end{equation}

For the positive frequency eigenfunctions we find the following final
expression%
\begin{equation}
\psi _{\sigma }^{(+)}(x)=\eta ^{2}C^{(+)}e^{i\left( qj\phi +kz\right)
}\left(
\begin{array}{c}
H_{1/2-im\alpha }^{(1)}(\gamma \eta )J_{\beta _{1}}(\lambda r)e^{-iq\phi /2}
\\
C_{\varphi }^{(1)}H_{1/2-im\alpha }^{(1)}(\gamma \eta )J_{\beta
_{2}}(\lambda r)e^{iq\phi /2} \\
-isH_{-1/2-im\alpha }^{(1)}(\gamma \eta )J_{\beta _{1}}(\lambda r)e^{-iq\phi
/2} \\
-isC_{\varphi }^{(1)}H_{-1/2-im\alpha }^{(1)}(\gamma \eta )J_{\beta
_{2}}(\lambda r)e^{iq\phi /2}%
\end{array}%
\right) ,  \label{psisig+1}
\end{equation}%
where $C^{(+)}=C_{1\varphi }^{(1)}$ and $C_{\varphi }^{(1)}$ is defined by (%
\ref{sval}). Recall that, in this formula $\eta =\alpha e^{-t/\alpha }$. The
coefficient $C^{(+)}$ in (\ref{psisig+1}) is determined from the
orthonormalization condition%
\begin{equation}
\int d^{3}x\sqrt{|g|}\psi _{\sigma }^{(-)+}\psi _{\sigma ^{\prime
}}^{(-)}=\delta _{\sigma \sigma ^{\prime }},  \label{norm}
\end{equation}%
where $g$ is the determinant of the metric tensor corresponding to the line
element (\ref{ds21}). The delta symbol on the rhs of (\ref{norm}) is
understood as the Kronecker delta for the discrete indices ($j$, $s$) and as
the Dirac delta function for the continuous ones ($\lambda $, $k$). By using
the Wronskian for the Hankel functions we find%
\begin{equation}
|C^{(+)}|^{2}=\frac{q\lambda e^{m\alpha \pi }}{32\pi \alpha ^{3}}(\gamma
+sk).  \label{C+}
\end{equation}

In a similar way, for the negative frequency eigenfunctions corresponding to
the Bunch-Davies vacuum state we find the following expression%
\begin{equation}
\psi _{\sigma }^{(-)}(x)=C^{(-)}\eta ^{2}e^{-i\left( qj\phi +kz\right)
}\left(
\begin{array}{c}
H_{1/2-im\alpha }^{(2)}(\gamma \eta )J_{\beta _{2}}(\lambda r)e^{-iq\phi /2}
\\
C_{\varphi }^{(2)}H_{1/2-im\alpha }^{(2)}(\gamma \eta )J_{\beta
_{1}}(\lambda r)e^{iq\phi /2} \\
-isH_{-1/2-im\alpha }^{(2)}(\gamma \eta )J_{\beta _{2}}(\lambda r)e^{-iq\phi
/2} \\
-isC_{\varphi }^{(2)}H_{-1/2-im\alpha }^{(2)}(\gamma \eta )J_{\beta
_{1}}(\lambda r)e^{iq\phi /2}%
\end{array}%
\right) ,  \label{psisig-}
\end{equation}%
with $s=\pm 1$ and%
\begin{equation}
C_{\varphi }^{(2)}=\frac{i\epsilon _{j}\lambda }{k-s\gamma },\;|C^{(-)}|^{2}=%
\frac{q\lambda e^{-m\alpha \pi }}{32\pi \alpha ^{3}}(\gamma -sk).  \label{C2}
\end{equation}%
For eigenfunctions (\ref{psisig-}) one has%
\begin{equation}
\widehat{J}_{3}\psi _{\sigma }^{(-)}=-qj\psi _{\sigma }^{(-)},\;j=\pm
1/2,\pm 3/2,\ldots .  \label{J3min}
\end{equation}%
The complete set of solutions described above may be used for the
investigation of field-theoretical effects induced by the conical structure
of the spacetime.

It is well-known that in the case of a scalar field in
$(D+1)$-dimensional dS spacetime the Bunch-Davies vacuum state is
not a physically realizable state for $m^{2}\alpha ^{2}\leqslant
-D(D+1)\xi $, where $\xi $ is the curvature coupling parameter. In
particular, this is the case for a minimally coupled massless
field. The corresponding Wightman function contains infrared
divergences arising from long-wavelength modes. For a fermionic
field, the Bunch-Davies vacuum state is physically realizable
independent of the mass.

\section{Fermionic condensate}

\label{sec:Cond}

In this section we evaluate the fermionic condensate assuming that the field
is prepared in the Bunch-Davies vacuum state. Having the complete set of
eigenfunctions we can evaluate the corresponding vacuum expectation value
(VEV) by using the mode-sum formula%
\begin{equation}
\langle 0|\bar{\psi}\psi |0\rangle =\sum_{\sigma }\bar{\psi}_{\sigma
}^{(-)}(x)\psi _{\sigma }^{(-)}(x),  \label{FermCond}
\end{equation}%
where
\begin{equation}
\sum_{\sigma }=\int_{-\infty }^{+\infty }dk\int_{0}^{\infty }d\lambda
\sum_{j=\pm 1/2,\pm 3/2,\ldots }\sum_{s=\pm 1}.  \label{Sumsig}
\end{equation}%
Of course, the expression on the right of (\ref{FermCond}) is divergent and
some renormalization procedure is needed. The important point here is that
for points outside the string the local geometry of dS spacetime is not
changed by the presence of the cosmic string. As a result the divergences
and the renormalization procedure are the same as those in dS spacetime when
the string is absent. In what follows, we implicitly assume the presence of
a cutoff function which makes the mode-sum finite. By taking into account
expression (\ref{psisig-}) for the negative frequency fermionic
eigenfunctions and introducing instead of the Hankel functions the MacDonald
function $K_{\nu }(x)$, after some transformations we find the following
expression%
\begin{eqnarray}
\langle 0|\bar{\psi}\psi |0\rangle  &=&\frac{q\eta ^{4}}{\pi ^{3}\alpha ^{3}}%
\sum_{j}\int_{0}^{\infty }dk\int_{0}^{\infty }d\lambda \,\lambda \gamma %
\left[ J_{qj+1/2}^{2}(\lambda r)+J_{qj-1/2}^{2}(\lambda r)\right]   \notag \\
&&\times \left[ |K_{1/2-im\alpha }(i\gamma \eta )|^{2}-|K_{1/2+im\alpha
}(i\gamma \eta )|^{2}\right] ,  \label{FC1}
\end{eqnarray}%
where for the summation over $j$ we have $\sum_{j}=\sum_{j=1/2,3/2,\ldots }$.

For the further transformation of this expression we use the formula%
\begin{equation}
|K_{1/2-im\alpha }(ix)|^{2}-|K_{1/2+im\alpha }(ix)|^{2}=-i\left( \partial
_{x}+\frac{1-2im\alpha }{x}\right) K_{1/2-im\alpha }(ix)K_{1/2-im\alpha
}(-ix),  \label{Krel}
\end{equation}%
which is easily obtained on the basis of the well-known properties of the
MacDonald function. This allows to present the fermionic condensate in the
form%
\begin{eqnarray}
\langle 0|\bar{\psi}\psi |0\rangle &=&\frac{q\eta ^{3}}{i\pi ^{3}\alpha ^{3}}%
\left( \eta \partial _{\eta }+1-2im\alpha \right) \sum_{j}\int_{0}^{\infty
}dk\int_{0}^{\infty }d\lambda \,\lambda  \notag \\
&&\times \left[ J_{qj+1/2}^{2}(\lambda r)+J_{qj-1/2}^{2}(\lambda r)\right]
K_{1/2-im\alpha }(i\gamma \eta )K_{1/2-im\alpha }(-i\gamma \eta ).
\label{FC2}
\end{eqnarray}%
For the separate integrals with the Bessel functions we use the formula (\ref%
{J4}) from appendix \ref{sec:App1} with the result%
\begin{eqnarray}
\langle 0|\bar{\psi}\psi |0\rangle &=&\frac{q}{(2\pi )^{3/2}i\pi }\left(
\frac{\eta }{\alpha r}\right) ^{3}\left( \eta \partial _{\eta }+1-2im\alpha
\right) \sum_{j}\int_{0}^{\infty }dx\,x^{1/2}  \notag \\
&&\times e^{x(\eta ^{2}/r^{2}-1)}\left[ I_{qj+1/2}(x)+I_{qj-1/2}(x)\right]
K_{1/2-im\alpha \nu }(x\eta ^{2}/r^{2}).  \label{FC3}
\end{eqnarray}%
By using the relation%
\begin{equation}
\left( u\partial _{u}+2\nu \right) e^{u^{2}}K_{\nu }(u^{2})=2u^{2}e^{u^{2}}
\left[ K_{\nu }(u^{2})-K_{\nu -1}(u^{2})\right] ,  \label{Krel1}
\end{equation}%
formula (\ref{FC3}) can also be written in the form%
\begin{eqnarray}
\langle 0|\bar{\psi}\psi |0\rangle &=&\frac{8q\alpha ^{-3}}{(2\pi )^{5/2}}%
\sum_{j}\int_{0}^{\infty }dy\,y^{3/2}e^{y(1-r^{2}/\eta ^{2})}{\mathrm{Im}}%
\left[ K_{1/2-im\alpha }(y)\right]  \notag \\
&&\times \left[ I_{qj+1/2}(yr^{2}/\eta ^{2})+I_{qj-1/2}(yr^{2}/\eta ^{2})%
\right] .  \label{FC4}
\end{eqnarray}%
In particular, from this formula it follows that the fermionic condensate
vanishes for a massless field. Hence, as in the case of a cosmic string in
flat spacetime, the presence of the string does not break chiral symmetry of
the massless theory. (In a previous publication \cite{Beze08}, we have
analyzed quantum fermionic fields in cosmic string spacetime obeying MIT bag
condition on a cylindrical boundary. There we have shown that the chiral
symmetry is automatically broken for massless fields.)

The expectation value (\ref{FC4}) contains both contributions coming from
the curvature of the de Sitter spacetime and from the non-trivial topology
induced by the cosmic string. In order to extract explicitly the part
induced by the string we consider the difference%
\begin{equation}
\langle \bar{\psi}\psi \rangle _{\text{s}}=\langle 0|\bar{\psi}\psi
|0\rangle -\langle 0|\bar{\psi}\psi |0\rangle _{\text{dS}},  \label{FCst}
\end{equation}%
where $\langle 0|\bar{\psi}\psi |0\rangle _{\text{dS}}$ is the fermionic
condensate in dS spacetime in the absence of the cosmic string. The formal
expression for the latter is obtained from (\ref{FC4}) taking $q=1$. The
corresponding renormalization procedure using the cutoff function is
described in Appendix \ref{sec:App2} and the renormalized fermionic
condensate is given by expression (\ref{FCdSren}). Due to the maximal
symmetry of dS spacetime and the dS invariance of the Bunch-Davies vacuum
state, the corresponding VEV\ does not depend on the spacetime point. It
vanishes for a massless field and is negative in the case of massive
fermionic field.

As it was mentioned above, for points away form the string axis, the
divergences do not depend on the presence of the cosmic string. Hence, the
string-induced part, $\langle \bar{\psi}\psi \rangle _{\text{s}}$, is finite
outside the string. In order to investigate this quantity we need to
evaluate the difference $\sum_{j}\left[ qI_{qj\pm 1/2}(x)-I_{j\pm 1/2}(x)%
\right] $. A convenient expression for this difference is provided by using
the Abel-Plana summation formula in the form (see, \cite{Most97,SahaRev})
\begin{equation}
\sum_{j}f(j)=\int_{0}^{\infty }du\,f(u)-i\int_{0}^{\infty }du\,\frac{%
f(iu)-f(-iu)}{e^{2\pi u}+1}.  \label{AbPl}
\end{equation}%
From here it follows that%
\begin{equation}
\sum_{j}\left[ qf(qj)-f(j)\right] =-i\int_{0}^{\infty }du\,\left[
f(iu)-f(-iu)\right] \left( \frac{1}{e^{2\pi u/q}+1}-\frac{1}{e^{2\pi u}+1}%
\right) .  \label{AbPl2}
\end{equation}%
Applying this formula for the series under consideration, one finds the
following integral representation:%
\begin{equation}
\sum_{j}\left[ qI_{qj+1/2}(x)+qI_{qj-1/2}(x)-I_{j+1/2}(x)-I_{j-1/2}(x)\right]
=\frac{4}{\pi }\int_{0}^{\infty }du\,g(q,u){\mathrm{Im}}[K_{1/2-iu}(x)],
\label{jSum}
\end{equation}%
where the notation%
\begin{equation}
g(q,u)=\cosh (u\pi )\left( \frac{1}{e^{2\pi u/q}+1}-\frac{1}{e^{2\pi u}+1}%
\right) ,  \label{gqx}
\end{equation}%
is introduced.

By making use of formula (\ref{jSum}), for the correction in the fermionic
condensate due to the presence of the string we find the expression%
\begin{eqnarray}
\langle \bar{\psi}\psi \rangle _{\text{s}} &=&\frac{4\sqrt{2}}{\pi
^{7/2}\alpha ^{3}}\int_{0}^{\infty }du\,g(q,u)\,\int_{0}^{\infty
}dy\,y^{3/2}e^{y(1-r^{2}/\eta ^{2})}  \notag \\
&&\times {\mathrm{Im}}[K_{1/2-iu}(yr^{2}/\eta ^{2})]\,{\mathrm{Im}}\left[
K_{1/2-im\alpha }(y)\right] .  \label{FCst2}
\end{eqnarray}%
The rhs of this formula is finite for points outside the string axis and the
cutoff function, implicitly assumed before, may be safely removed. As it is
seen from (\ref{FCst2}), the part in the fermionic condensate induced by the
string is a function of the radial and time coordinates in the form of the
ratio $r/\eta $. This property is a consequence of the maximal symmetry of
dS spacetime and the Bunch-Davies vacuum state. Note that the proper
distance from the axis of the string is given by $\alpha r/\eta $. Hence,
the ratio $r/\eta $ is the proper distance from the string measured in units
of the dS curvature radius $\alpha $.

Let us consider the asymptotic behavior of the fermionic condensate at small
and large distances from the string. Introducing a new integration variable $%
x=yr^{2}/\eta ^{2}$ in (\ref{FCst2}), we see that for points near the string
the argument of the MacDonald function with the order $1/2-im\alpha $ is
large. Taking into account that for large values $\ y$ and for fixed $%
m\alpha $ one has%
\begin{equation}
{\mathrm{Im}}[K_{1/2-im\alpha }(y)]\sim -\frac{m\alpha }{2y}\sqrt{\frac{\pi
}{2y}}e^{-y},  \label{Klarge}
\end{equation}%
we find
\begin{equation}
\langle \bar{\psi}\psi \rangle _{\text{s}}\approx -\frac{2m\eta ^{2}}{\pi
^{3}\alpha ^{2}r^{2}}\int_{0}^{\infty }du\,g(q,u)\,\int_{0}^{\infty
}dx\,e^{-x}{\mathrm{Im}}[K_{1/2-iu}(x)].  \label{FClarge01}
\end{equation}%
The integral over $x$ is evaluated with the help of the formula \cite{Prud86}%
\begin{equation}
\int_{0}^{\infty }dx\,x^{\beta -1}e^{-x}K_{\nu }(x)=\frac{\sqrt{\pi }}{%
2^{\beta }}\frac{\Gamma (\beta +\nu )\Gamma (\beta -\nu )}{\Gamma (\beta
+1/2)}.  \label{IntFormK1}
\end{equation}%
After the integration over $u$ this leads to the final result%
\begin{equation}
\langle \bar{\psi}\psi \rangle _{\text{s}}\approx \frac{m(q^{2}-1)}{24\pi
^{2}(\alpha r/\eta )^{2}},\;r/\eta \ll 1.  \label{FCsmall}
\end{equation}%
Hence, on the string axis the condensate diverges as the inverse second
power of the proper distance. The part in the fermionic condensate
corresponding to dS spacetime without string is constant everywhere and,
hence, near the string, the condensate is dominated by the string-induced
part.

In the opposite limit of large distances from the string, the main
contribution to the integral over $y$ in (\ref{FCst2}) comes from the region
near the lower limit of the integration. By using the asymptotic formula for
the MacDonald function for small values of the argument, to the leading
order, we find%
\begin{eqnarray}
\langle \bar{\psi}\psi \rangle _{\text{s}} &\approx &\frac{4(\eta /r)^{4}}{%
\pi ^{7/2}\alpha ^{3}}\int_{0}^{\infty }du\,g(q,u)\,{\mathrm{Im}}\bigg[%
\Gamma (1/2-im\alpha )  \notag \\
&&\times \left( \frac{\eta ^{2}}{2r^{2}}\right) ^{im\alpha }\int_{0}^{\infty
}dx\,x^{1+im\alpha }e^{-x}{\mathrm{Im}}[K_{1/2-iu}(x)]\bigg].
\label{FClarge0}
\end{eqnarray}%
By taking into account that $2i{\mathrm{Im}}%
[K_{1/2-iu}(x)]=K_{1/2-iu}(x)-K_{1/2+iu}(x)$, the integral over $x$ is
evaluated with the help of formula (\ref{IntFormK1}). As a result, the
leading term in the asymptotic expansion of the fermionic condensate is
presented in the form%
\begin{equation}
\langle \bar{\psi}\psi \rangle _{\text{s}}\approx \frac{\alpha A(q,m\alpha )%
}{\pi ^{3}(\alpha r/\eta )^{4}}\,\sin \left[ 2m\alpha \ln (2r/\eta )-\varphi
_{0}\right] ,\;r/\eta \gg 1,  \label{FClarge}
\end{equation}%
where the function $A(q,m\alpha )\geqslant 0$ and the phase $\varphi _{0}$
are defined by the relation%
\begin{equation}
A(q,m\alpha )e^{i\varphi _{0}}=\frac{\Gamma (1/2-im\alpha )}{\Gamma
(5/2+im\alpha )}\int_{0}^{\infty }du\,ug(q,u)\Gamma (3/2+im\alpha -iu)\Gamma
(3/2+im\alpha +iu).  \label{AqRel}
\end{equation}%
Hence, at large distances the behavior of the string induced part in the
fermionic condensate is damping oscillatory with the amplitude decaying as
the inverse fourth power of the proper distance. The value of the ratio $%
r/\eta $ corresponding to the first zero of the condensate increases with
decreasing $m\alpha $. The same is the case for the distance between the
neighbor zeros. Note that in the case of a scalar field, in dependence of
the curvature radius of dS spacetime, at large distances two regimes are
realized with monotonic and oscillatory behavior of the VEVs for the field
squared and the energy-momentum tensor \cite{Beze09}. In deriving asymptotic
formula (\ref{FClarge}) we have assumed that the value of $m\alpha $ is
fixed and, hence, $m\alpha r/\eta \gg 1$. The latter means that the proper
distance from the string axis is much larger than the Compton length of the
spinor particle. So, at these distances we have a power-law suppression of
the string-induced VEV. This is in contrast to the case of flat spacetime,
in which at distances larger than the Compton length one has an exponential
suppression \cite{Beze06}.

Now let us show that in the limit $\alpha \rightarrow \infty $ and for fixed
$t$, from formula (\ref{FCst2}) the expression of the fermionic condensate
is obtained for the geometry of a cosmic string in Minkowski spacetime. In
this limit one has $\eta \approx \alpha $. In formula (\ref{FCst2}) we
introduce a new integration variable $x=yr^{2}/\alpha ^{2}$. In the
corresponding integral over $x$ the main contribution comes from the region
with $x\lesssim 1$ and for a fixed value of $r$ the argument of the
MacDonald function with the order $1/2-im\alpha $ is large. The leading term
in the asymptotic expansion of this function can be obtained by making use
of the integral representation $K_{\nu }(z)=\int_{0}^{\infty }dt\,e^{-z\cosh
t}\cosh (\nu t)$. For the imaginary part entering in (\ref{FCst2}), for
fixed $z$ and $m$ and for large values $\alpha $ we find%
\begin{equation}
{\mathrm{Im}}\left[ K_{1/2-im\alpha }(z\alpha ^{2})\right] \approx -\frac{m%
\sqrt{\pi }}{(2z)^{3/2}\alpha ^{2}}e^{-m^{2}/2z-z\alpha ^{2}}.  \label{ImKas}
\end{equation}%
Using this asymptotic formula, from (\ref{FCst2}), in the limit $\alpha
\rightarrow \infty $, for the fermionic condensate we find%
\begin{equation}
\langle \bar{\psi}\psi \rangle _{\text{s}}\rightarrow \langle \bar{\psi}\psi
\rangle _{\text{s}}^{\text{(M)}}=-\frac{2m}{\pi ^{3}r^{2}}\int_{0}^{\infty
}du\,g(q,u)\,\int_{0}^{\infty }dx\,{\mathrm{Im}}%
[K_{1/2-iu}(x)]e^{-m^{2}r^{2}/2x-x}.  \label{FCMink}
\end{equation}%
The expression in the rhs of this formula coincides with the integral
representation of the fermionic condensate for the string in Minkowski
spacetime found in \cite{Beze08}.

\begin{figure}[tbph]
\begin{center}
\begin{tabular}{cc}
\epsfig{figure=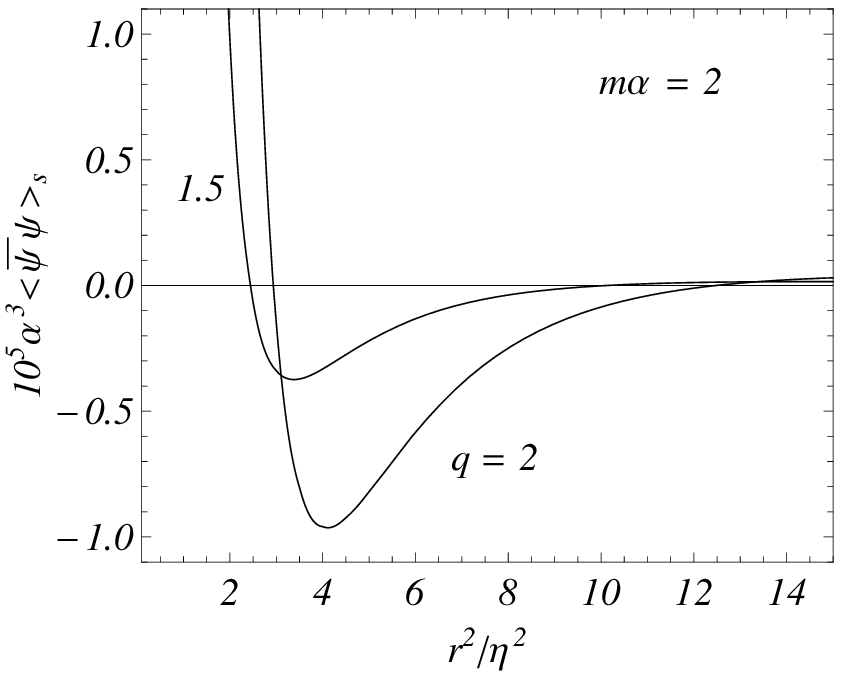,width=7.cm,height=6.cm} & \quad %
\epsfig{figure=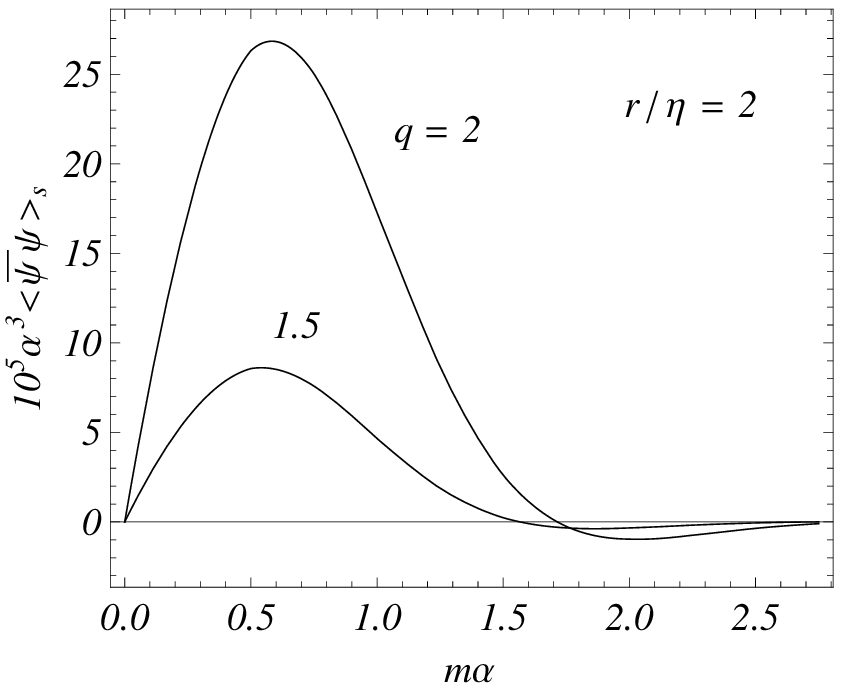,width=7.cm,height=6.cm}%
\end{tabular}%
\end{center}
\caption{String-induced part in the fermionic condensate as a function of
the ratio $r^{2}/\protect\eta ^{2}$ (left plot) and of the field mass (right
plot.)}
\label{fig1}
\end{figure}

In the left plot of figure \ref{fig1} we have presented the dependence of
the string-induced part in the fermionic condensate as a function of the
ratio $r^{2}/\eta ^{2}$ for fixed values $q=1.5,2$ and for $m\alpha =2$. The
right plot presents the dependence of the same quantity as a function of the
field mass for $r/\eta =2$. Recall that the ratio $r/\eta $ is the proper
distance from the string measured in units of the dS curvature radius $%
\alpha $.

\section{Energy-momentum tensor}

\label{sec:EMT}

Another important quantity which characterizes the properties of the quantum
vacuum is the VEV\ of the energy-momentum tensor. In addition to describing
the physical structure of the quantum field at a given point, the
energy-momentum tensor acts as a source of gravity in the Einstein equations
and plays an important role in modelling self-consistent dynamics involving
the gravitational field. In this section we consider the VEV of the
energy-momentum tensor for the geometry of a cosmic string in dS spacetime.
The corresponding VEV for a fermionic field in the geometry of a cosmic
string in flat spacetime is investigated in \cite%
{Frol87,Dowk87b,Line95,More95,Beze06}. In particular, the VEV for a massive
Dirac field is considered in \cite{Beze06}. In a recent paper \cite{Beze08},
the VEV of the energy-momentum tensor is analyzed for a massive spinor field
obeying the MIT bag boundary condition on a cylindrical shell in the cosmic
string spacetime.

The VEV of the energy-momentum tensor for the fermionic field can be
evaluated by making use of the mode-sum formula
\begin{equation}
\langle 0|T_{\mu \nu }|0\rangle =\frac{i}{2}\sum_{\sigma }[\bar{\psi}%
_{\sigma }^{(-)}(x)\gamma _{(\mu }\nabla _{\nu )}\psi _{\sigma
}^{(-)}(x)-(\nabla _{(\mu }\bar{\psi}_{\sigma }^{(-)}(x))\gamma _{\nu )}\psi
_{\sigma }^{(-)}(x)]\ ,  \label{modesum}
\end{equation}%
with the negative frequency eigenspinors given by (\ref{psisig-}). The
brackets in the index expressions mean the symmetrization over the indices
enclosed. Note that, as before, we are working in the coordinate system $%
(t,r,\phi ,z)$. As in the case of the fermionic condensate, the presence of
the cutoff function is assumed in the rhs of (\ref{modesum}). Taking into
account expressions (\ref{Gaml}) for the components of the spin connection,
we see that (no summation) $\{\gamma _{\mu },\Gamma _{\mu }\}=0$, where the
figure braces stand for the anticommutator. From here it follows that the
terms in (\ref{modesum}) with the spin connection will not contribute to the
VEVs of diagonal components.

First we consider the vacuum energy density. Substituting the expression for
the eigenfunctions from (\ref{psisig-}), after long but straightforward
calculations, the corresponding mode-sum can be presented in the form%
\begin{eqnarray}
\langle 0|T_{0}^{0}|0\rangle  &=&-\frac{q\eta ^{5}}{2\pi ^{3}\alpha ^{4}}%
\sum_{j}\int_{0}^{+\infty }dk\int_{0}^{\infty }d\lambda \,\lambda \left[
J_{qj+1/2}^{2}(\lambda r)+J_{qj-1/2}^{2}(\lambda r)\right]   \notag \\
&\times &\left[ \partial _{\eta }^{2}+\frac{2}{\eta }\partial _{\eta
}+4\left( \gamma ^{2}+im\alpha \frac{1/2-im\alpha }{\eta ^{2}}\right) \right]
K_{1/2-im\alpha }(i\gamma \eta )K_{1/2-im\alpha }(-i\gamma \eta ).
\label{T00}
\end{eqnarray}%
The rhs of this formula is expressed in terms of the functions $\mathcal{J}%
_{\beta }(r,\eta )$ and $\mathcal{I}_{\beta }(r,\eta )$ given in appendix %
\ref{sec:App1}. By using the corresponding formulae we find%
\begin{eqnarray}
\langle 0|T_{0}^{0}|0\rangle  &=&-\frac{q(\eta /r)^{5}}{2^{1/2}\pi
^{5/2}\alpha ^{4}}\sum_{j}\int_{0}^{\infty }dx\,x^{3/2}e^{-x}  \notag \\
&&\times \left[ I_{qj+1/2}(x)+I_{qj-1/2}(x)\right] \hat{F}%
_{u}e^{u}K_{1/2-im\alpha }(u)|_{u=x\eta ^{2}/r^{2}},  \label{T002}
\end{eqnarray}%
where we have defined the operator%
\begin{equation}
\hat{F}_{u}=u\partial _{u}^{2}+\left( 3/2-2u\right) \partial _{u}-2+im\alpha
\left( 1/2-im\alpha \right) /u.  \label{Fhat}
\end{equation}%
Making use the properties of the MacDonald function it can be seen that%
\begin{equation}
\hat{F}_{u}e^{u}K_{1/2-im\alpha }(u)=-\frac{1}{2}e^{u}\left[ K_{1/2-im\alpha
}(u)+K_{-1/2-im\alpha }(u)\right] .  \label{FhatK}
\end{equation}%
This leads to the following final result for the VEV\ of the energy density%
\begin{eqnarray}
\langle 0|T_{0}^{0}|0\rangle  &=&\frac{4q\alpha ^{-4}}{(2\pi )^{5/2}}%
\sum_{j}\int_{0}^{\infty }dy\,y^{3/2}e^{y(1-r^{2}/\eta ^{2})}{\mathrm{Re}}%
\left[ K_{1/2-im\alpha }(y)\right]   \notag \\
&&\times \left[ I_{qj+1/2}(yr^{2}/\eta ^{2})+I_{qj-1/2}(yr^{2}/\eta ^{2})%
\right] .  \label{T004}
\end{eqnarray}%
As in the case of the fermionic condensate, this VEV is a function of the
ratio $r/\eta $.

Now let us consider the vacuum stress along the axis of the cosmic string.
From the mode-sum formula (\ref{modesum}) for this component we have%
\begin{equation}
\langle 0|T_{33}|0\rangle =(\alpha /\tau )\sum_{\sigma }k\psi _{\sigma
}^{(-)+}\gamma ^{(0)}\gamma ^{(3)}\psi _{\sigma }^{(-)}.  \label{T33}
\end{equation}%
After the substitution of the expressions for the eigenfunctions, this leads
to the result%
\begin{eqnarray}
\langle 0|T_{3}^{3}|0\rangle  &=&\frac{q\eta ^{5}}{\pi ^{3}\alpha ^{4}}%
\sum_{j}\int_{0}^{\infty }dk\,k^{2}\int_{0}^{\infty }d\lambda \lambda \left[
J_{qj+1/2}^{2}(\lambda r)+J_{qj-1/2}^{2}(\lambda r)\right]   \notag \\
&\times &\left[ K_{1/2-im\alpha }(i\gamma \eta )K_{1/2-im\alpha }(-i\gamma
\eta )+K_{1/2+im\alpha }(i\gamma \eta )K_{1/2+im\alpha }(-i\gamma \eta )%
\right] .  \label{T332}
\end{eqnarray}%
Using the integral representation (\ref{IntRep}) for the products of the
MacDonald functions and the procedure similar to that described in Appendix %
\ref{sec:App1}, it can be seen that%
\begin{equation}
\langle 0|T_{3}^{3}|0\rangle =\langle 0|T_{0}^{0}|0\rangle .  \label{T333}
\end{equation}%
Although this result is expected in a pure cosmic string spacetime due boost
invariance of the spacetime along the $z$-direction, this is not the case
for the present geometry. In fact, for a scalar field, the equality (\ref%
{T333}), in general, is not obeyed \cite{Beze09}.

For the radial stress, from (\ref{modesum}), taking into account expression (%
\ref{psisig-}), after long transformations we can see that%
\begin{eqnarray}
\langle 0|T_{1}^{1}|0\rangle  &=&\frac{q\eta ^{5}}{\pi ^{3}\alpha ^{4}}%
\sum_{j}\int_{0}^{\infty }dk\int_{0}^{\infty }d\lambda \,\lambda ^{3}\left[
J_{qj-1/2}(\lambda r)J_{qj+1/2}^{\prime }(\lambda r)-J_{qj-1/2}^{\prime
}(\lambda r)J_{qj+1/2}(\lambda r)\right]   \notag \\
&&\times \left[ K_{1/2-im\alpha }(i\gamma \eta )K_{1/2-im\alpha }(-i\gamma
\eta )+K_{1/2+im\alpha }(i\gamma \eta )K_{1/2+im\alpha }(-i\gamma \eta )%
\right] .  \label{T112}
\end{eqnarray}%
For the further transformation we use the relation%
\begin{eqnarray}
&&J_{qj-1/2}(\lambda r)J_{qj+1/2}^{\prime }(\lambda r)-J_{qj-1/2}^{\prime
}(\lambda r)J_{qj+1/2}(\lambda r)  \notag \\
&&\quad =\frac{1}{\lambda ^{2}}\left( \frac{1}{2}\partial _{r}^{2}+\frac{1}{r%
}\partial _{r}-qj\frac{2qj-1}{r^{2}}\right) J_{qj-1/2}^{2}(\lambda
r)+2J_{qj-1/2}^{2}(\lambda r),  \label{relBess}
\end{eqnarray}%
for the Bessel function. After the changing of the order of integrations and
differentiation, the part in the VEV with the first term on the rhs of (\ref%
{relBess}) is evaluated in the way similar to that for the axial stress. The
part with the second term on the right of (\ref{relBess}) is evaluated
similar to that described in Appendix \ref{sec:App1} for the function $%
\mathcal{J}_{\beta }(r,\eta )$ with the only difference that now we need the
integral $\int_{0}^{\infty }d\lambda \,\lambda ^{3}J_{\beta }^{2}(\lambda
r)e^{-p\lambda ^{2}}$. The formula for the latter is obtained from (\ref%
{IntAp}) differentiating with respect to $p$. Combining the results of these
calculations, we obtain the following result%
\begin{equation}
\langle 0|T_{1}^{1}|0\rangle =\frac{q(\eta /r)^{5}}{2^{1/2}\pi ^{5/2}\alpha
^{4}}\sum_{j}\int_{0}^{\infty }dx\,x^{3/2}e^{x\eta ^{2}/r^{2}}{\mathrm{Re}}%
\left[ K_{1/2-im\alpha }(x\eta ^{2}/r^{2})\right] \hat{G}%
_{x}e^{-x}I_{qj-1/2}(x),  \label{T113}
\end{equation}%
with the operator%
\begin{equation}
\hat{G}_{x}=2x\partial _{x}^{2}+(4x+3)\partial _{x}+4-qj\frac{2qj-1}{x}.
\label{Gop}
\end{equation}%
Now, by using the properties of the modified Bessel function, it can be seen
that%
\begin{equation}
\hat{G}_{x}e^{-x}I_{qj-1/2}(x)=e^{-x}\left[ I_{qj+1/2}(x)+I_{qj-1/2}(x)%
\right] .  \label{GopI}
\end{equation}%
Hence, we see that the radial stress is equal to the energy density:%
\begin{equation}
\langle 0|T_{1}^{1}|0\rangle =\langle 0|T_{0}^{0}|0\rangle .  \label{T11T00}
\end{equation}%
Again, in general, this property is not the case for a scalar field.

Now we turn to the azimuthal stress. The substitution of the negative
frequency eigenfunctions from (\ref{psisig-}) to the corresponding mode-sum
leads to the result%
\begin{eqnarray}
\langle 0|T_{2}^{2}|0\rangle  &=&\frac{2q^{2}\eta ^{5}}{\pi ^{3}\alpha ^{4}r}%
\sum_{j}j\int_{0}^{\infty }dk\int_{0}^{\infty }d\lambda \,\lambda
^{2}J_{qj-1/2}(\lambda r)J_{qj+1/2}(\lambda r)  \notag \\
&\times &\left[ K_{1/2-im\alpha }(i\gamma \eta )K_{1/2-im\alpha }(-i\gamma
\eta )+K_{1/2+im\alpha }(i\gamma \eta )K_{1/2+im\alpha }(-i\gamma \eta )%
\right] .  \label{T222}
\end{eqnarray}%
For the further transformation, first, we use the relation%
\begin{equation}
\lambda J_{qj-1/2}(\lambda r)J_{qj+1/2}(\lambda r)=\left( \frac{qj-1/2}{r}-%
\frac{1}{2}\partial _{r}\right) J_{qj-1/2}^{2}(\lambda r).  \label{relJ}
\end{equation}%
Changing the order of integrations and the operator on the rhs of (\ref{relJ}%
), the integrals can be transformed in the way similar to that used before.
We come to the following result%
\begin{eqnarray}
\langle 0|T_{2}^{2}|0\rangle  &=&\frac{\sqrt{2}q^{2}}{\pi ^{5/2}\alpha ^{4}}%
\sum_{j}j\int_{0}^{\infty }dy\,y^{3/2}e^{y}{\mathrm{Re}}\left[
K_{1/2-im\alpha }(y)\right]   \notag \\
&&\times \left( \frac{qj-1/2}{x}-\partial _{x}\right)
e^{-x}I_{qj-1/2}(x)|_{x=yr^{2}/\eta ^{2}}.  \label{T223}
\end{eqnarray}%
Taking into account that
\begin{equation}
\left( \frac{qj-1/2}{x}-\partial _{x}\right)
e^{-x}I_{qj-1/2}(x)=e^{-x}\left[
I_{qj-1/2}(x)-I_{qj+1/2}(x)\right] ,  \label{relI}
\end{equation}%
we finally get%
\begin{eqnarray}
\langle 0|T_{2}^{2}|0\rangle  &=&\frac{8q^{2}}{(2\pi )^{5/2}\alpha ^{4}}%
\sum_{j}j\int_{0}^{\infty }dy\,y^{3/2}e^{y(1-r^{2}/\eta ^{2})}{\mathrm{Re}}%
\left[ K_{1/2-im\alpha }(y)\right]   \notag \\
&&\times \left[ I_{qj-1/2}(yr^{2}/\eta ^{2})-I_{qj+1/2}(yr^{2}/\eta ^{2})%
\right] .  \label{T224}
\end{eqnarray}

It remains to consider the off-diagonal component $\langle 0|T_{01}|0\rangle
$ (other components vanish by the symmetry of the problem). The
corresponding mode-sum has the form%
\begin{equation}
\langle 0|T_{01}|0\rangle =\frac{i}{4}\sum_{\sigma }[\psi _{\sigma
}^{(-)+}\partial _{1}\psi _{\sigma }^{(-)}-(\partial _{1}\psi _{\sigma
}^{(-)+})\psi _{\sigma }^{(-)}+\psi _{\sigma }^{(-)+}\gamma ^{0}\gamma
_{1}\partial _{0}\psi _{\sigma }^{(-)}-(\partial _{0}\psi _{\sigma
}^{(-)+})\gamma ^{0}\gamma _{1}\psi _{\sigma }^{(-)}].  \label{T01mode}
\end{equation}%
First of all, by using the expression for the functions $\psi _{\sigma
}^{(-)}$ it can be seen that $\psi _{\sigma }^{(-)+}\partial _{1}\psi
_{\sigma }^{(-)}=(\partial _{1}\psi _{\sigma }^{(-)+})\psi _{\sigma }^{(-)}$%
. Second, the expression of the last two terms on the rhs of (\ref{T01mode})
contain the factor $C_{\varphi }^{(2)}+C_{\varphi }^{(2)\ast }$ which
vanishes in accordance with (\ref{C2}). Hence, we conclude that the
off-diagonal component $\langle 0|T_{01}|0\rangle $ vanishes. At this point
we would like to say that for the case of a scalar field this component, in
general, is non-zero \cite{Beze09}. It vanishes only for a conformally
coupled massless field.

Similar to the case of the fermionic condensate, we define the subtracted
VEV of the energy-momentum tensor:%
\begin{equation}
\langle T_{\mu }^{\nu }\rangle _{\text{s}}=\langle 0|T_{\mu }^{\nu
}|0\rangle -\langle 0|T_{\mu }^{\nu }|0\rangle _{\text{dS}},  \label{SubTik}
\end{equation}%
where $\langle 0|T_{\mu }^{\nu }|0\rangle _{\text{dS}}$ is the corresponding
VEV in dS spacetime when the cosmic string is absent. The expressions for
the components of the latter are obtained from the formulae given above in
this section taking $q=1$. The corresponding renormalization procedure based
on the introduction of the cutoff function is described in Appendix \ref%
{sec:App2}. The renormalized VEV of the energy-momentum tensor in dS
spacetime without the string is given by (\ref{TlkdS1}) and it has been
previously derived in \cite{Mama81} (see also \cite{Grib94}) using the $n$%
-wave regularization method.

The explicit expressions for the string-induced parts in the energy density,
axial and radial stresses are obtained using formula (\ref{jSum}) (no
summation over $\mu $):%
\begin{eqnarray}
\langle T_{\mu }^{\mu }\rangle _{\text{s}} &=&\frac{4\alpha ^{-4}}{%
2^{1/2}\pi ^{7/2}}\int_{0}^{\infty }dx\,g(q,x)\int_{0}^{\infty
}dy\,y^{3/2}e^{y(1-r^{2}/\eta ^{2})}\,  \notag \\
&&\times {\mathrm{Im}}[K_{1/2-ix}(yr^{2}/\eta ^{2})]{\mathrm{Re}}\left[
K_{1/2-im\alpha }(y)\right] ,  \label{Tlls}
\end{eqnarray}%
with $\mu =0,1,3$. In order to obtain the expression for the string-induced
part in the azimuthal stress we employ the summation formula%
\begin{eqnarray}
&&\sum_{j}j\left[
q^{2}I_{qj+1/2}(u)-I_{j+1/2}(u)-q^{2}I_{qj-1/2}(u)+I_{j-1/2}(u)\right]
\notag \\
&&\quad =-\frac{4}{\pi }\int_{0}^{\infty }dx\,xg(q,x){\mathrm{Re}}\left[
K_{1/2+ix}(u)\right] ,  \label{SForm}
\end{eqnarray}%
which follows from the Abel-Plana formula (\ref{AbPl}). This leads to the
following expression%
\begin{eqnarray}
\langle T_{2}^{2}\rangle _{\text{s}} &=&\frac{8\alpha ^{-4}}{2^{1/2}\pi
^{7/2}}\int_{0}^{\infty }dx\,xg(q,x)\int_{0}^{\infty
}dy\,y^{3/2}e^{y(1-r^{2}/\eta ^{2})}  \notag \\
&&\times {\mathrm{Re}}\left[ K_{1/2-ix}(yr^{2}/\eta ^{2})\right] {\mathrm{Re}%
}\left[ K_{1/2-im\alpha }(y)\right] .  \label{T22s}
\end{eqnarray}%
The string-induced parts (\ref{Tlls}) and (\ref{T22s}) are finite at points
outside the string and the renormalization is needed only for the dS part
without string.

Now let us check that the string-induced parts in the VEVs obey the trace
relation%
\begin{equation}
\langle T_{\mu }^{\mu }\rangle _{\text{s}}=m\langle \bar{\psi}\psi \rangle _{%
\text{s}}.  \label{TraceRel}
\end{equation}%
In order to see this, we write the function in the integrand of the trace in
the form%
\begin{equation}
3{\mathrm{Im}}[K_{1/2-ix}(u)]+2x{\mathrm{Re}}\left[ K_{1/2-ix}(u)\right]
=2(1-u){\mathrm{Im}}[K_{1/2-ix}(u)]+2\partial _{u}{\mathrm{Im}}\left[
uK_{1/2-ix}(u)\right] .  \label{TrRel}
\end{equation}%
with $u=yr^{2}/\tau ^{2}$. Then, in the part with the last term on the right
of (\ref{TrRel}) we use integration by parts. In particular, from (\ref%
{TraceRel}) it follows that the string-induced part in the VEV\ of the
energy-momentum tensor is traceless for a massless fermionic field. The
trace anomaly is contained in the pure dS part. In addition to (\ref%
{TraceRel}), the string-induced parts obey the covariant conservation
equation $\nabla _{\nu }\langle T_{\mu }^{\nu }\rangle _{\text{s}}=0$. By
taking into account that $\langle T_{\mu }^{\nu }\rangle _{\text{s}}$ is a
function of the ratio $r/\eta $, this equation reduces to the relation%
\begin{equation}
\langle T_{2}^{2}\rangle _{\text{s}}=\partial _{u}(u\langle T_{0}^{0}\rangle
_{\text{s}}),\;u=r/\eta .  \label{ConsEq}
\end{equation}%
This relation is easily checked making use of the formula $\left( \partial
_{u}u\pm 2qj\right) e^{-yu^{2}}I_{qj\pm 1/2}(yu^{2})=0$ for the modified
Bessel function.

The general formulae for the string-induced parts in the VEVs are simplified
for a special case of a massless field. Noting that $K_{1/2}(y)=\sqrt{\pi /2y%
}e^{-y}$, for the integral over $y$ we use the formula%
\begin{equation}
\int_{0}^{\infty }dy\,ye^{-yr^{2}/\eta ^{2}}K_{1/2-ix}(yr^{2}/\eta ^{2})=%
\frac{\pi (4x^{2}+1)}{24\cosh (\pi x)}\frac{3+2ix}{(r/\eta )^{4}}.
\label{IntFormK}
\end{equation}%
After the evaluation of the remained integral over $x$, we get the result%
\begin{equation}
\langle T_{\mu }^{\nu }\rangle _{\text{s}}=-\frac{(q^{2}-1)(7q^{2}+17)}{%
2880\pi ^{2}(\alpha r/\eta )^{4}}\text{diag}(1,1,-3,1).  \label{Tklm0}
\end{equation}%
The massless fermionic field is conformally invariant and we could obtain
this formula directly by conformal transformation of the corresponding
result for the cosmic string in Minkowski spacetime \cite{Frol87}. Note
that, in the same way, by taking into account that the electromagnetic field
is conformally invariant in four-dimensional spacetime, for the
corresponding string-induced part in dS spacetime, by using the result from
\cite{Frol87}, we find%
\begin{equation}
\langle T_{\mu }^{\nu }\rangle _{\text{s}}^{\text{(el)}}=-\frac{%
(q^{2}-1)(q^{2}+11)}{720\pi ^{2}(\alpha r/\eta )^{4}}\text{diag}(1,1,-3,1).
\label{TklEl}
\end{equation}%
Hence, with this paper we complete the investigations of the string-induced
vacuum polarization for scalar, fermionic and electromagnetic fields in dS
spacetime.

For a massive fermionic field the formulae for the string-induced parts in
the VEV of the energy-momentum tensor are simplified at small and large
distances from the cosmic string. For points near the string the main
contribution to the integrals in (\ref{Tlls}) comes from large values of $y$%
. For these $y$ one has $K_{1/2-im\alpha }(y)\approx K_{1/2}(y)$ and, hence,
near the string, to the leading order, the string-induced parts in the VEV
of the energy-momentum tensor coincide with the corresponding expressions
for a massless field, given by (\ref{Tklm0}). At large distances from the
string, the main contributions to the integrals over $y$ come from small
values of $y$. In a way similar to that used for the case of the fermionic
condensate, to the leading order we find (no summation over $\mu $)%
\begin{equation}
\langle T_{\mu }^{\mu }\rangle _{\text{s}}\approx -\frac{A(q,m\alpha )}{2\pi
^{3}(\alpha r/\eta )^{4}}\cos \left[ 2m\alpha \ln (2r/\eta )-\varphi _{0}%
\right] ,  \label{Tlllarge}
\end{equation}%
for $\mu =0,1,3$, and
\begin{equation}
\langle T_{2}^{2}\rangle _{\text{s}}\approx \frac{A(q,m\alpha )}{\pi
^{3}(\alpha r/\eta )^{4}}\sqrt{9/4+m^{2}\alpha ^{2}}\sin [2m\alpha \ln
(2r/\eta )-\varphi _{0}+\varphi _{1}],  \label{T22large}
\end{equation}%
where $\varphi _{1}=\arctan [3/(2m\alpha )]$. Here, the function $%
A(q,m\alpha )$ and the phase $\varphi _{0}$ are defined by formula (\ref%
{AqRel}). Now the trace relation between the asymptotics of the
energy-momentum tensor and the fermionic condensate is explicitly observed.
Note that the oscillations in the energy density and in the fermionic
condensate are shifted by the phase $\pi /2$.

As in the case of the fermionic condensate, we can check that in the limit $%
\alpha \rightarrow \infty $ for fixed values of $t$, $r$ and $m$, the VEV of
the energy-momentum tensor in the geometry of a string in Minkowski
spacetime is obtained. In a way similar to that used for (\ref{ImKas}), it
can be seen that in this limit we have%
\begin{equation}
{\mathrm{Re}}\left[ K_{1/2-im\alpha }(z\alpha ^{2})\right] \approx \frac{1}{%
\alpha }\sqrt{\frac{\pi }{2x}}e^{-m^{2}/2z-z\alpha ^{2}}.  \label{ReKas}
\end{equation}%
Now, with the help of this formula, in the limit under consideration one
finds (no summation)%
\begin{eqnarray}
\langle T_{\mu }^{\mu }\rangle _{\text{s}} &\rightarrow &\langle T_{\mu
}^{\mu }\rangle _{\text{s}}^{\text{(M)}}=\frac{2}{\pi ^{3}r^{4}}%
\int_{0}^{\infty }du\,g(q,u)\int_{0}^{\infty }dx\,x{\mathrm{Im}}%
[K_{1/2-iu}(x)]e^{-m^{2}r^{2}/2x-x},  \notag \\
\langle T_{2}^{2}\rangle _{\text{s}} &\rightarrow &\langle T_{2}^{2}\rangle
_{\text{s}}^{\text{(M)}}=\frac{4}{\pi ^{3}r^{4}}\int_{0}^{\infty
}du\,ug(q,u)\int_{0}^{\infty }dx\,x{\mathrm{Re}}\left[ K_{1/2-iu}(x)\right]
e^{-m^{2}r^{2}/2x-x},  \label{T22Mink}
\end{eqnarray}%
with $\mu =0,1,3$. Again, these expressions coincide with the integral
representations given in \cite{Beze08} (with the missprint in the expression
for the energy density corrected).

\begin{figure}[tbph]
\begin{center}
\begin{tabular}{cc}
\epsfig{figure=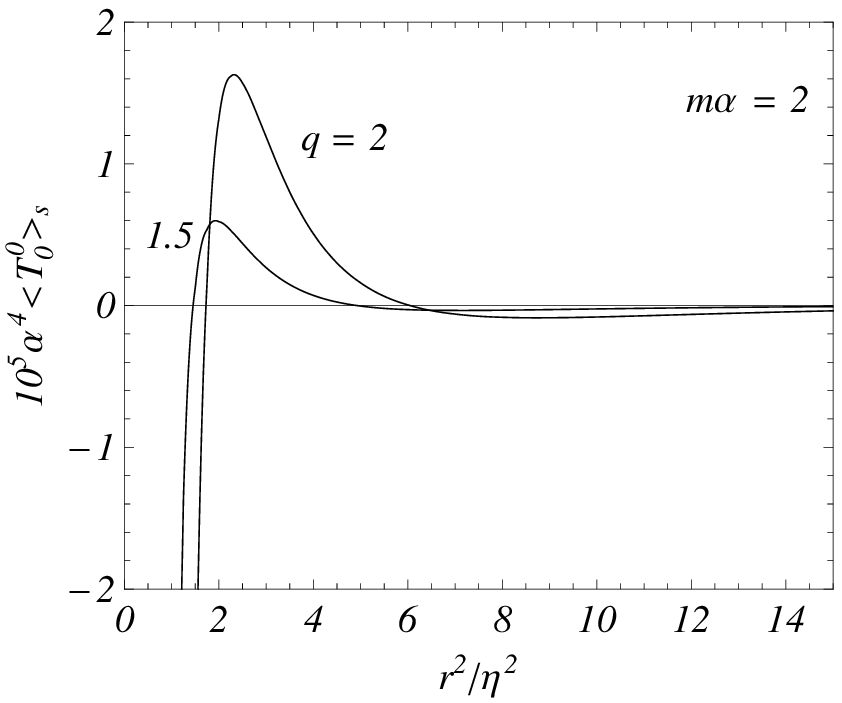,width=7.cm,height=6.cm} & \quad %
\epsfig{figure=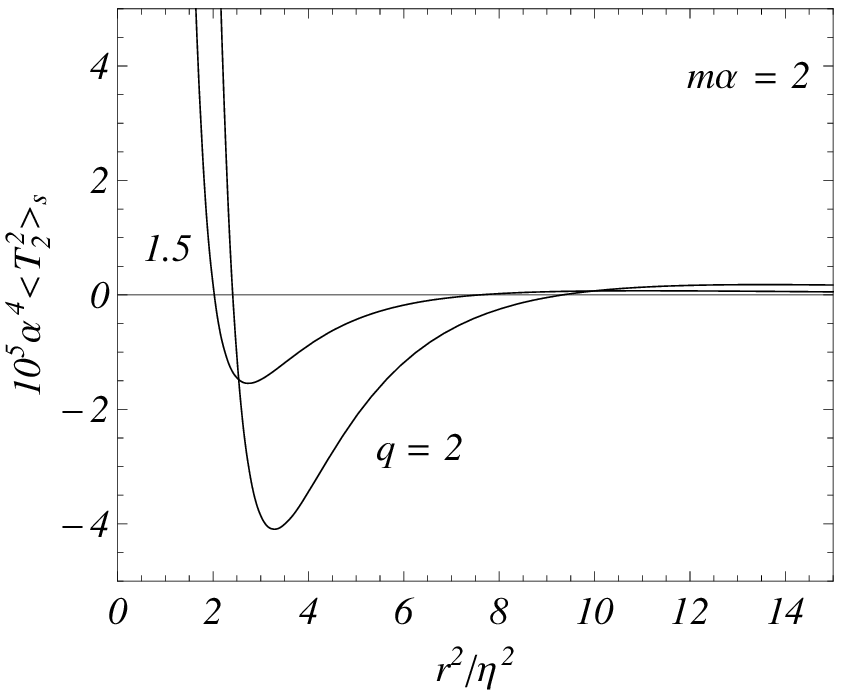,width=7.cm,height=6.cm}%
\end{tabular}%
\end{center}
\caption{String-induced parts in the energy density (left plot) and the
azimuthal stress (right plot) as functions of the ratio $r^{2}/\protect\eta %
^{2}$ for fixed values $q=1.5,2$ and for $m\protect\alpha =2$.}
\label{fig2}
\end{figure}

\begin{figure}[tbph]
\begin{center}
\begin{tabular}{cc}
\epsfig{figure=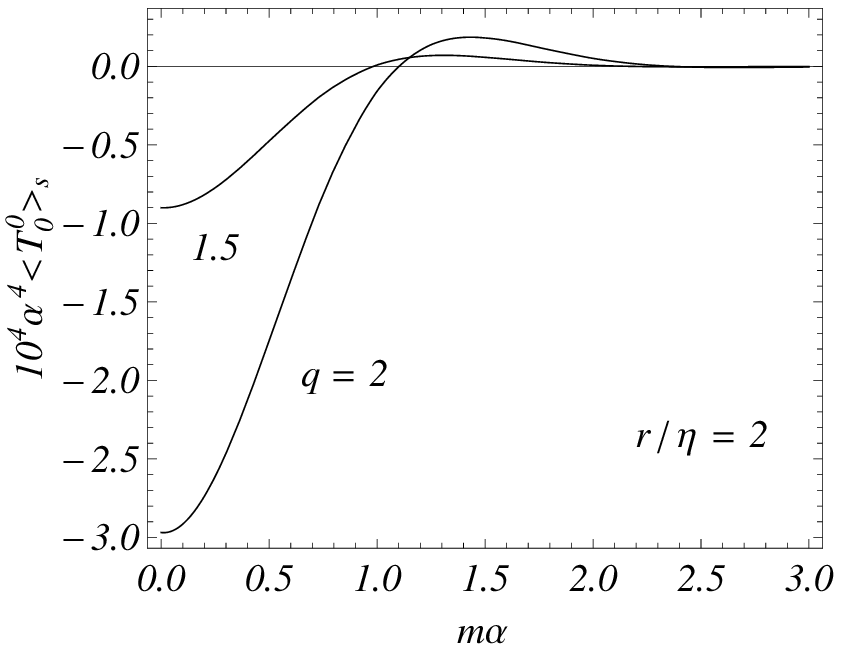,width=7.cm,height=6.cm} & \quad %
\epsfig{figure=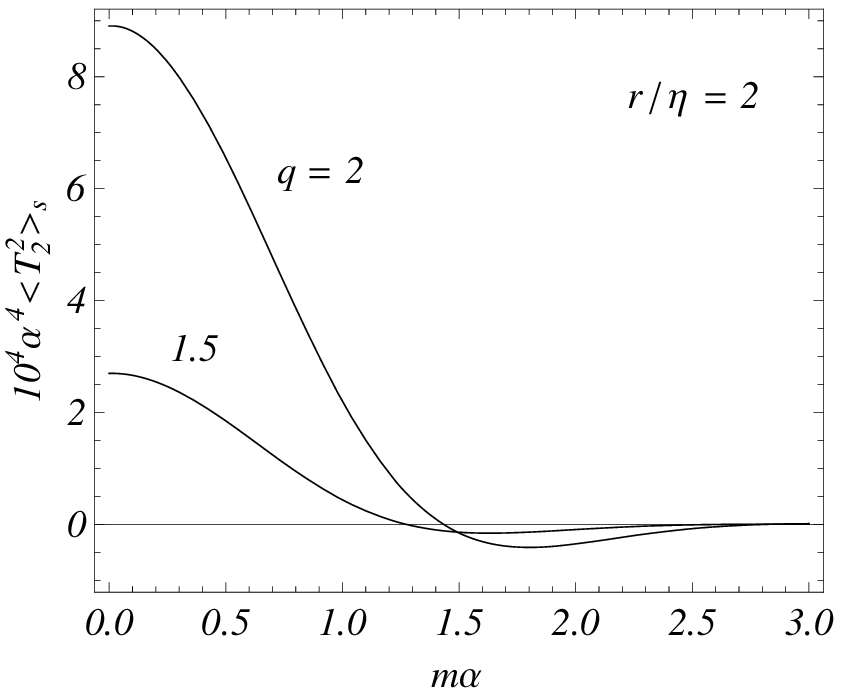,width=7.cm,height=6.cm}%
\end{tabular}%
\end{center}
\caption{String-induced parts in the energy density (left plot) and the
azimuthal stress (right plot) as functions of the field mass for $r/\protect%
\eta =2$.}
\label{fig3}
\end{figure}

In figure \ref{fig2} we have plotted the string induced parts in the VEVs of
the energy-momentum tensor components as functions of $r^{2}/\eta ^{2}$ for
fixed values $q=1.5,2$ and for $m\alpha =2$. The left/right plot corresponds
to the energy density/azimuthal stress. Figure \ref{fig3} presents the same
quantities as functions of $m\alpha $ for $r/\eta =2$.

\section{Conclusion}

\label{sec:Conc}

In the present paper we have investigated the polarization of the fermionic
vacuum induced by a cosmic string in dS spacetime. We have assumed that the
field is prepared in the Bunch-Davies vacuum state. Unlike to the case of a
scalar field, for a fermionic field the Bunch-Davies vacuum is a physically
realizable state independent of the field mass. Among the most important
covariant characteristics of the vacuum state are the fermionic condensate
and the expectation value of the energy-momentum tensor. In order to
evaluate these quantities we have used the direct mode-summation method. In
this approach the complete set of normalized eigenfunctions for the Dirac
equation is used. In section \ref{sec:EigFunc} we have found these
eigenfunction for the geometry under consideration. They are given by
expressions (\ref{psisig+1}) and (\ref{psisig-}) for the positive and
negative frequency solutions, respectively.

On the basis of these expressions, in section \ref{sec:Cond} we have
investigated the fermionic condensate. The corresponding mode-sum is given
by expression (\ref{FC4}). This expression contains both contributions
coming from the curvature of the de Sitter spacetime and from the
non-trivial topology induced by the cosmic string. In order to extract
explicitly the part induced by the string, we have subtracted the term
corresponding to the geometry of dS spacetime when the string is absent. Due
to the maximal symmetry of dS spacetime and the dS invariance of the
Bunch-Davies vacuum state, the corresponding VEV\ does not depend on the
spacetime point. Using the Abel-Plana summation formula, the string-induced
part in the fermionic condensate is presented in the form (\ref{FCst2}).
This part vanishes for a massless field. As the presence of the cosmic
string does not change the local geometry outside the string axis, the
renormalization in the calculation of the corresponding condensate is
reduced to that for dS spacetime without the string.

The general formula for the string-induced part in the fermionic condensate
is simplified in asymptotic regions of small and large distances from the
axis of the string. For points near the string the leading term in the
corresponding asymptotic expansion is given by expression (\ref{FCsmall})
and the condensate diverges as the inverse square of the proper distance.
The part in the fermionic condensate corresponding to dS spacetime without
string is constant everywhere and near the string, the condensate is
dominated by the string-induced part. At large distances from the string the
behavior of the string-induced part in the fermionic condensate is described
by formula (\ref{FClarge}) and this behavior is damping oscillatory with the
amplitude decaying with the inverse fourth power of the distance. The value
of the proper distance corresponding to the first zero of the condensate
increases with decreasing $m\alpha $. The same is the case for the distance
between the neighbor zeros. As an additional check, we have shown that in
the limit of large dS curvature radius the result is recovered for the
fermionic condensate in the geometry of a cosmic string on background of
flat spacetime.

The VEV of the energy-momentum tensor for a fermionic field is investigated
in section \ref{sec:EMT}. Unlike to the case of a scalar field, this VEV is
diagonal. The axial and radial stresses are equal to the energy density. The
corresponding mode-sums are transformed to the form (\ref{T004}), for the
energy density, and to the form (\ref{T224}), for the azimuthal stress.
Similar to the case of the fermionic condensate, we have extracted from the
VEVs the parts corresponding to dS spacetime without the string. For the
subtracted part, integral representations are obtained useful for the
numerical evaluation. For a massless field the string-induced part in the
VEV of the energy-momentum tensor is given by a simple formula (\ref{Tklm0})
and is related to the corresponding flat spacetime result by the standard
conformal transformation. In this case the energy density and stresses are
monotonic functions of the distance from the string. This is not the case
for a massive field. The corresponding VEVs are given by formulae (\ref{Tlls}%
) and (\ref{T22s}). These VEVs depend on the radial and time coordinates in
the combination $r/\eta $ which is the proper distance from the string in
units of the dS curvature radius. For points near the cosmic string, the
leading term in the corresponding asymptotic expansion does not depend on
the field mass and the string-induced part diverges as the inverse fourth
power of the distance. At large distances, the behavior of the
string-induced parts is damping oscillatory. As for the case of the
condensate, the amplitude of these oscillations decays with the inverse
fourth power of the distance. The oscillations in the energy density and in
the fermionic condensate are shifted by the phase $\pi /2$. The power-law
decay of the string-induced VEVs at large distances from the string is in
contrast to the case of a string in flat spacetime, in which at distances
larger than the Compton length of a spinor particle one has an exponential
suppression.

The presence of a cosmic string in the period of inflation will
lead to the modification of the power-spectrum for vacuum
fluctuations. The cosmic string breaks the homogeneity of dS space
and, as a result, the power-spectrum depends on the distance from
the string. In the case of a scalar field this dependence was
investigated in \cite{Saha10}, where it has been shown that the
influence of the string appears in the form of a universal
multiplier in the power-spectrum which does not depend on time, on
the field mass and on the curvature coupling parameter. An
interesting topic for the further investigation is the influence
of the fermionic fluctuations, discussed in the present paper, on
the power-spectrum from the inflation. The latter is imprinted in
the anisotropies of the cosmic microwave background radiation and
the observational data on these anisotropies could constrain the
density and the parameters of the cosmic strings at the end of
inflation.

The present paper completes the investigations of the
string-induced vacuum polarization for scalar, fermionic and
electromagnetic fields in dS spacetime. The case of scalar field
was considered in \cite{Beze09} and for the electromagnetic field
the corresponding VEV of the energy-momentum tensor is obtained
from the flat spacetime result by a simple conformal
transformation (see (\ref{TklEl})).

\section*{Acknowledgments}

E.R.B.M. thanks Conselho Nacional de Desenvolvimento Cient\'{\i}fico e Tecnol%
\'{o}gico (CNPq) and FAPES-ES/CNPq (PRONEX) for partial financial support.
A.A.S. was supported by Conselho Nacional de Desenvolvimento Cient\'{\i}fico
e Tecnol\'{o}gico (CNPq) and by the Armenian Ministry of Education and
Science Grant No. 119.

\appendix

\section{Integral representation}

\label{sec:App1}

In this appendix we consider the transformation of the expression%
\begin{equation}
\mathcal{J}_{\beta }(r,\eta )=\int_{0}^{\infty }dk\int_{0}^{\infty }d\lambda
\lambda J_{\beta }^{2}(\lambda r)K_{\nu }(i\gamma \eta )K_{\nu }(-i\gamma
\eta ).  \label{J}
\end{equation}%
As the first step we use the following integral representation for the
product of the MacDonald functions \cite{Wats44}:%
\begin{equation}
K_{\nu }(i\gamma \eta )K_{\nu }(-i\gamma \eta )=\int_{0}^{\infty
}du\,u^{-1}\int_{0}^{\infty }dy\,\cosh (2\nu y)\exp [u(\gamma \eta
)^{2}-(\gamma \eta )^{2}u\cosh (2y)-1/2u].  \label{IntRep}
\end{equation}%
Substituting this in (\ref{J}), the integral over $k$ is taken explicitly:%
\begin{eqnarray}
\mathcal{J}_{\beta }(r,\eta ) &=&\frac{\sqrt{\pi }}{2^{3/2}\eta }%
\int_{0}^{\infty }d\lambda \lambda J_{\beta }^{2}(\lambda r)\int_{0}^{\infty
}du\,u^{-3/2}\int_{0}^{\infty }dy\,  \notag \\
&&\times \frac{\cosh (2\nu y)}{\sinh y}\exp [-2u(\lambda \eta )^{2}\sinh
^{2}y-1/2u].  \label{J2}
\end{eqnarray}%
As the next step, for the $\lambda $-integral we use the formula \cite%
{Prud86}%
\begin{equation}
\int_{0}^{\infty }d\lambda \,\lambda J_{\beta }^{2}(\lambda r)e^{-p\lambda
^{2}}=\frac{e^{-r^{2}/2p}}{2p}I_{\beta }(r^{2}/2p).  \label{IntAp}
\end{equation}%
After introducing a new integration variable $x=r^{2}/(4u\eta ^{2}\sinh
^{2}y)$, we obtain%
\begin{equation}
\mathcal{J}_{\beta }(r,\eta )=\frac{\pi ^{1/2}}{2^{1/2}r^{3}}%
\int_{0}^{\infty }dy\,\cosh (2\nu y)\int_{0}^{\infty
}dx\,x^{1/2}e^{-x}I_{\beta }(x)\exp \left( -2x\eta ^{2}\sinh
^{2}y/r^{2}\right) .  \label{J3}
\end{equation}%
Changing the order of integrations, the integral over $y$ is expressed in
terms of the MacDonald function and we obtain the following formula%
\begin{equation}
\mathcal{J}_{\beta }(r,\eta )=\frac{\sqrt{\pi }}{2^{3/2}r^{3}}%
\int_{0}^{\infty }dx\,x^{1/2}e^{x(\eta ^{2}/r^{2}-1)}I_{\beta }(x)K_{\nu
}(x\eta ^{2}/r^{2}).  \label{J4}
\end{equation}

Next, we consider the integral%
\begin{equation}
\mathcal{I}_{\beta }(r,\eta )=\int_{0}^{\infty }dk\int_{0}^{\infty }d\lambda
\lambda \gamma ^{2}J_{\beta }^{2}(\lambda r)K_{\nu }(i\gamma \eta )K_{\nu
}(-i\gamma \eta ).  \label{Ibet}
\end{equation}%
By taking into account (\ref{IntRep}), in a way similar to that for the
function $\mathcal{J}_{\beta }(r,\eta )$, it can be seen that%
\begin{equation}
\mathcal{I}_{\beta }(r,\eta )=-\frac{\sqrt{\pi /2}}{r^{5}}\partial _{\eta
^{2}}\eta ^{2}\int_{0}^{\infty }dx\,x^{3/2}e^{x(\eta ^{2}/r^{2}-1)}I_{\beta
}(x)K_{\nu }(x\eta ^{2}/r^{2}).  \label{Ibet2}
\end{equation}%
Formulae (\ref{J4}) and (\ref{Ibet2}) are used in the main text in order to
derive integral representations for string-induced parts in the fermionic
condensate and in the VEV\ of the energy-momentum tensor.

\section{Fermionic vacuum densities in dS spacetime without string}

\label{sec:App2}

In this section we consider the VEVs in dS spacetime when the string is
absent. The corresponding energy-momentum tensor is investigated in \cite%
{Mama81} by using the $n$-wave regularization method. Here we show
that our approach based on the cutoff function method leads to the
same result. In addition, we will evaluate the fermionic
condensate.

First we consider the fermionic condensate. The corresponding integral
representation is obtained from (\ref{FC4}) by taking $q=1$. In this case
the series over $j$ is summed by using the formula $I_{0}(x)+2\sum_{n=1}^{%
\infty }I_{n}(x)=e^{x}$. Introducing explicitly the exponential
cutoff, we find%
\begin{equation}
\langle 0|\bar{\psi}\psi |0\rangle _{\text{dS}}^{(\beta )}=\frac{8\alpha
^{-3}}{(2\pi )^{5/2}}\int_{0}^{\infty }dy\,y^{3/2}e^{(1-\beta )y}{\mathrm{%
Im\,}}\left[ K_{1/2-im\alpha }(y)\right] ,  \label{dScut}
\end{equation}%
where $\beta >0$ is a cutoff parameter. The integral in this formula is
expressed in terms of the associated Legendre function of the first kind
(see \cite{Prud86}). We will write the corresponding result in terms of the
hypergeometric function:%
\begin{equation}
\int_{0}^{\infty }dy\,y^{3/2}e^{(1-\beta )y}K_{1/2-im\alpha }(y)=-\frac{i\pi
\sqrt{\pi /2}}{\sinh (\pi m\alpha )}\partial _{\beta }^{2}F(im\alpha
,-im\alpha +1;1;1-\beta /2).  \label{IntCut}
\end{equation}%
A convenient expansion in powers of $\beta $ for the
hypergeometric function on the right hand side of this formula is
given in \cite{Abra64}. For the
fermionic condensate this leads to the expansion%
\begin{equation}
\langle 0|\bar{\psi}\psi |0\rangle _{\text{dS}}^{(\beta )}=-\frac{m\alpha }{%
4\pi ^{2}\alpha ^{3}}\left\{ \frac{2}{\beta }+(1+m^{2}\alpha ^{2})\left[ \ln
(\beta /2)+2{\mathrm{Re\,}}\psi (im\alpha )-2\ln (ma)+b_{1}\right] +o(\beta
)\right\} ,  \label{FCcut}
\end{equation}%
where $\psi (x)$ is the digamma function. The presence of the logarithmic
term in this expansion signifies about the renormalization non uniqueness in
the form of $b_{1}$. This non uniqueness can be removed by imposing an
additional renormalization condition. As such a condition we will require
that $\left\langle \bar{\psi}\psi \right\rangle _{\text{dS,ren}}\rightarrow
0 $ in the limit $m\rightarrow \infty $ (for the discussion of this
condition in the context of the Casimir effect see \cite{Bord97}). With this
condition for the renormalized fermionic condensate in dS spacetime we find%
\begin{equation}
\langle \bar{\psi}\psi \rangle _{\text{dS,ren}}=\frac{m}{2\pi ^{2}\alpha ^{2}%
}\left[ (1+m^{2}\alpha ^{2})\left[ \ln (m\alpha )-{\mathrm{Re\,}}\psi
(im\alpha )\right] +1/12\right] .  \label{FCdSren}
\end{equation}%
It can be checked that expression (\ref{FCdSren}) coincides with the result
obtained from the effective Lagrangian with the help of formula $\langle
\bar{\psi}\psi \rangle _{\text{dS,ren}}=-(2/\sqrt{|g|})\partial _{m}L_{\text{%
eff}}$. The expression for the effective Lagrangian in dS spacetime is
derived in \cite{Cand75} by using the dimensional regularization. For large
masses, $m\alpha \gg 1$, in the leading order we have%
\begin{equation}
\langle \bar{\psi}\psi \rangle _{\text{dS,ren}}\approx -\frac{11}{480\pi
^{2}\alpha ^{4}m}.  \label{FCdSlarge}
\end{equation}%
For a massless field the fermionic condensate in dS spacetime vanishes. In
the limt $m\alpha \ll 1$, for the leading term one has $\langle \bar{\psi}%
\psi \rangle _{\text{dS,ren}}\approx m\ln (m\alpha )/(2\pi ^{2}\alpha ^{2})$%
. The numerical evaluation shows that $\langle \bar{\psi}\psi \rangle _{%
\text{dS,ren}}<0$ for a massive fermionic field with the minimum \ at $%
m\alpha \approx 0.27$.

Now we turn to the energy-momentum tensor. Taking $q=1$ in formulae (\ref%
{T004}) and (\ref{T224}) we find
\begin{equation}
\langle 0|T_{\mu }^{\nu }|0\rangle _{\text{dS}}^{(\beta
)}=\frac{4\alpha ^{-4}\delta _{\mu }^{\nu }}{(2\pi
)^{5/2}}\int_{0}^{\infty }dy\,y^{3/2}e^{(1-\beta
)y}{\mathrm{Re\,}}\left[ K_{1/2-im\alpha }(y)\right] .
\label{TlkdS}
\end{equation}%
As we could expect from the maximal symmetry of dS spacetime, all
components coincide. By using formula (\ref{IntCut}), expanding
the hypergeometric function and imposing the renormalization
condition $\langle T_{\mu }^{\nu }\rangle _{\text{ren}}\rightarrow
0$ for $m\rightarrow \infty $, for the
renormalized VEV of the energy-momentum tensor we find%
\begin{equation}
\langle T_{\mu }^{\nu }\rangle _{\text{ren}}=\frac{\alpha ^{-4}\delta _{\mu
}^{\nu }}{8\pi ^{2}}\left\{ m^{2}\alpha ^{2}(1+m^{2}\alpha ^{2})\left[ \ln
(m\alpha )-{\mathrm{Re\,}}\psi (im\alpha )\right] +\frac{m^{2}\alpha ^{2}}{12%
}+\frac{11}{120}\right\} .  \label{TlkdS1}
\end{equation}%
This expression coincides with the one derived in \cite{Mama81} by
using the $n$-wave regularization method. In the massless limit,
Eq. (\ref{TlkdS1}) gives the well-known expression for the trace
anomaly. The energy density corresponding to (\ref{TlkdS1}) has
the maximum for a massless field, becomes zero at $m\alpha \approx
0.467$ and is negative for larger values of this parameter. For
large values of the field mass, $m\alpha \gg 1$, to the leading
order we have
\begin{equation}
\langle T_{\mu }^{\nu }\rangle _{\text{ren}}\approx -\frac{\delta _{\mu
}^{\nu }\alpha ^{-6}}{960\pi ^{2}m^{2}},  \label{TlkdSLarge}
\end{equation}%
and the VEV vanishes, in accordance with renormalization condition
imposed.

\end{document}